\documentclass[12pt]{article}
\usepackage{mathrsfs}
\usepackage{amssymb}
\usepackage{dsfont}
\usepackage{amsmath}
\usepackage{graphicx}
\usepackage{subfigure}
\usepackage{etex}

\normalsize
\input{epsf}

\makeatletter

\newcommand{\Rmnum}[1]{\expandafter\@slowromancap\romannumeral #1@}
\makeatother

\begin{document}
\addtolength{\baselineskip}{.20mm}
\newlength{\extraspace}
\setlength{\extraspace}{2mm}
\newlength{\extraspaces}
\setlength{\extraspaces}{2mm}

\newcommand{\newsection}[1]{
\vspace{15mm} \pagebreak[3] \addtocounter{section}{1}
\setcounter{subsection}{0} \setcounter{footnote}{0}
\noindent {\Large\bf \thesection. #1} \nopagebreak
\medskip
\nopagebreak}
\newcommand{\newsubsection}[1]{
\vspace{1cm} \pagebreak[3] \addtocounter{subsection}{1}
\addcontentsline{toc}{subsection}{\protect
\numberline{\arabic{section}.\arabic{subsection}}{#1}}
\noindent{\large\bf 
\thesubsection. #1} \nopagebreak \vspace{3mm} \nopagebreak}
\newcommand{\ba}{\begin{eqnarray}
\addtolength{\abovedisplayskip}{\extraspaces}
\addtolength{\belowdisplayskip}{\extraspaces}

\addtolength{\belowdisplayshortskip}{\extraspace}}

\newcommand{\be}{\begin{equation}
\addtolength{\abovedisplayskip}{\extraspaces}
\addtolength{\belowdisplayskip}{\extraspaces}
\addtolength{\abovedisplayshortskip}{\extraspace}
\addtolength{\belowdisplayshortskip}{\extraspace}}
\newcommand{\ee}{\end{equation}}
\newcommand{\STr}{{\rm STr}}
\newcommand{\figuur}[3]{
\begin{figure}[t]\begin{center}
\leavevmode\hbox{\epsfxsize=#2 \epsffile{#1.eps}}\\[3mm]
\parbox{15.5cm}{\small
\it #3}
\end{center}
\end{figure}}
\newcommand{\im}{{\rm Im}}
\newcommand{\calm}{{\cal M}}
\newcommand{\call}{{\cal L}}
\newcommand{\sect}[1]{\section{#1}}
\newcommand\hi{{\rm i}}
\def\bea{\begin{eqnarray}}
\def\eea{\end{eqnarray}}

\begin{titlepage}
\begin{center}

\vspace{3.5cm}

{\Large \bf{Testing  the effects from dark radiation}}\\[1.5cm]

{Yi Zhang $^{a,b,c}$\footnote{Email: zhangyia@cqupt.edu.cn},}{Yungui Gong $^{c,d}$\footnote{Email: yggong@mail.hust.edu.cn}, } \vspace*{0.5cm}

{\it $^{a}$College of Mathematics and Physics, Chongqing University
of Posts and Telecommunications, \\ Chongqing 400065, China

 $^{b}$High Energy Physics Division, Argonne National Laboratory, \\ Lemont, IL 60439, USA}

 $^{c}$Institute of Theoretical Physics, Chinese Academy of Science,  \\ Beijing 100190, China

$^{d}$School of Physics, Huazhong University of Science and Technology, \\ Wuhan 430074, China
\date{\today}
\vspace{3.5cm}

\textbf{Abstract} \vspace{5mm}

 \end{center}
 In this letter, the effects of dark radiation (DR) are tested.  Theoretically, the phase-space analysis method is applied to
   check whether the model is consist with the history of our universe which shows  positive results.
    Observationally,  by using the observational data  ($SNLS$
   (SuperNovae Legacy Survey) ,   $WMAP9$(Wilkinson Microwave Anisotropy Probe 9 Years Result), $PLANCK$ (Planck First Data Release),
 $BAO$ (Baryon Acoustic Oscillations), $H(z)$ (Hubble Parameter Data) and  $BBN$ (Big Bang Nucleosynthesis)),
  the dark radiation is found to have the effect of  wiping out the tension between  the $SNLS$ data and the other data  in flat $\Lambda CDM $ model.
 The effects of dark radiation also make  the best fit value of $N_{eff}$  slightly larger than $3.04$.
\end{titlepage}

\section{Introduction}\label{sec1}

The observations hint our universe is accelerating now (e.g. Refs. \cite{Riess,Perlmutter99,Tonry03,Knop03,Riess04}). The observations also show a nearly flat universe with
roughly $72\%$ dark energy,  $28\%$ matter and $0.1\%$ radiation (e.g. Refs.\cite{Bennet03,spergel03,Page07,Hinshaw07,Komatsu:2010fb}).
 How to describe these observations by  theories? The   $\Lambda$CDM model is the simplest candidate.
In $\Lambda$CDM model,  the generation of  neutrino is assumed as three.   And, the number of the effective neutrino species is
 $N_{eff}=3.04$  where the effects from the
 non-instantaneous neutrino decoupling from the primordial photon-baryon plasma are taken into account.
However, many theoretical models indicate the existing of extra radiation,
   e.g. the FRW model in the
   Randall-Sundrum scenario  \cite{Randall:1999ee,Vishwakarma:2002ek,Ichiki:2002eh,Ichiki:2002yp};
    the Brans-Dicke theory \cite{Calabrese:2011hg,DeFelice:2005bx};
    the Horava-Lifshitz theory \cite{Dutta:2009jn,Ali:2010sv};
    the decaying vacuum \cite{Lima:2000ay,Birkel:1996py,Lima:1995kd};
     the negative Casimir effect \cite{Bordag:2001qi}.

    Recently, the measurement of the temperature anisotropy of CMB (Cosmic Microwave Background) shows less
power spectrum at small scale, suggesting that  $N_{eff}$ has a bigger value
than the one predicted by the standard model of particle physics, so
the existence of ``dark radiation". The results of   $WMAP7$ \cite{Komatsu:2010fb},
 $ACT$  (the Atacama Cosmology Telescope \cite{Dunkley:2010ge,ACT}) and $SPT$ (South Pole Telescope \cite{Keisler:2011aw}) 
  give out the $1\sigma$ level of the effective neutrino number are $N_{eff}= 4.56\pm0.75 $($WMAP7$),
  $N_{eff}= 2.78\pm0.55 $($WMAP7+ACT$), $2.96\pm0.44$ ($WMAP7+ACT+SPT$); while the BBN data shows $N_{eff}=3.24\pm 0.6$ \cite{Cyburt:2004yc}. 
  So many discussions on dark radiation have already appeared (e.g.  Refs.\cite{Fischler:2010xz,Zhang:2012zz,Hamann:2011hu,Menestrina:2011mz}).

In this letter,   the $\Lambda CDM $ 
model with dark radiation will be used to test the dark radiation effect which could be  generated from a  electroweak phase transition   \cite{Dutta:2009ix}. The letter is organized as follows. In Section \ref{sec2}, the model will be introduced.
  In Section \ref{3},  a phase-space analysis will be presented to get the evolution of our universe. Then  in Section \ref{sec4}, we apply the observation data to test the model parameter space, including
the $SNLS$ complication of  supernova Ia data
\cite{Conley:2011ku,Sullivan:2011kv}, the cosmic
  microwave background radiation data from $WMAP9$ \cite{Bennett:2012fp,Hinshaw:2012aka} and $PLANCK$ \cite{Ade:2013tyw},  the BAO distance measurements from the
 oscillations in the distribution of galaxies \cite{Percival:2007yw,Percival:2009xn,Blake:2011en,Beutler:2011hx},
 the Hubble parameter data \cite{Simon:2004tf,Riess:2009pu} and the BBN data \cite{Serra:2009yp,Burles:2000zk}.
    We will show
   the constraining results in Section \ref{sec5}.
 Finally, a short summary will be given out in  Section \ref{sec6}.

\section{The Model}\label{sec2}

 Here, the geometry of space-time is assumed to be described by the FRW (
Friedmann-Robertson-Walker) metric with a non-zero curvature,
\begin{eqnarray}
 ds^{2}=-dt^{2}+a^{2}(t)\left(\frac{dr^{2}}{1-kr^{2}}+r^{2}(d\theta^{2}+\sin ^{2}\theta
 d \varphi ^{2})\right),
\end{eqnarray}
where $a$ is the scale factor, and $k$ is the curvature parameter
with the values of  $0,\pm1$  representing flat, closed and open spatial sector
respectively.

The energy density components in our universe are represented by the pressureless matter part $\rho_m$, the dark energy
part $\rho_d$, the
ordinary radiation part $\rho_{r}$, the dark radiation part
$\rho_{dr}$ and the curvature part $\rho_k$. The Friedmann
equation is
\begin{eqnarray}
 H^{2}=\frac{1}{3m_{pl}^{2}}(\rho_{m}+\rho_{d}+\rho_{r}+\rho_{dr}+\rho_{k}),
\end{eqnarray}
where $\rho_{k}=-k/a^{2}$ and $m_{pl}$ is the Planck mass.  We call the  $\Lambda$CDM model plus the dark radiation  as the flat or curved dark radiation model.  This kind of model could be  derived  from a quintessence scenario phenomenologically
  whose   potential includes
interactions of the field with virtual particles and the heat bath. As the potential is
similar to the Higgs potential in the electroweak phase transition, a first-order phase
transition at redshift $z\sim3$ releases  energy in relativistic model (dark radiation).    
After that, $\rho_{d}$ becomes a constant; and the dark radiation appears \cite{Dutta:2009ix}.

After defining $\Omega_{i}=\rho_{i}/8\pi G H^{2} $, $\Omega_{m}$, $\Omega_{d}$, $\Omega_{r}$,  $\Omega_{dr}$ and $\Omega_{k}$  could 
represent the fractional energy
densities for matter, dark energy, ordinary radiation, dark radiation and curvature respectively. The energy components are assumed to be conserved separately. Specially,
\begin{eqnarray}
\Omega_{r}+\Omega_{dr}=\left[1+
\frac{7}{8}(\frac{4}{11})^{\frac{4}{3}}N_{eff}\right]\Omega_{\gamma},
\end{eqnarray}
where $h=H_{0}/100 Mpc.km.s^{-1}$,  the index ``0" denotes the present   value of  parameter and $\Omega_{\gamma}$ is the energy
density of the CMB photons background at temperature
$T_{\gamma}=2.728K$.
 To  represent the dark
radiation, we use the symbols $f=\Omega_{dr0}/\Omega_{d0}$   which represents the ratio of today's dark radiation and dark energy. Then, the Friedmann equation could be rewritten as below
 \begin{eqnarray}
 H^{2}=H_{0}^{2}[\Omega_{d0}+\Omega_{k0}a^{-2}+\Omega_{m0}a^{-3}+\Omega_{r0}a^{-4}
 +f\Omega_{d0}a^{-4}].
 \end{eqnarray}

If treating the dark radiation as a signal of dark energy, dark radiation leads to a characteristical time dependence in the effective  EoS (equation of state) parameter of dark energy,
 \begin{eqnarray}
 \label{omega}
\omega(z)= \frac{p_{d}+p_{dr}}{\rho_{d}+\rho_{dr}}=\frac{f(1+z)^{4}/3-1}{f(1+z)^{4}+1},
 \end{eqnarray}
where  $z$ is the redshift with the definition $z=a^{-1}-1$.
And, the time derivative of the EoS parameter  is
\begin{eqnarray}
\label{wprime}
\omega' =-\frac{16}{3}\frac{f(1+z)^{4}}{[f(1+z)^{4}+1]^{2}},
 \end{eqnarray}
 where a prime means the derivative with respect to $\ln a$.
Based on Eqs.\ref{omega} and \ref{wprime}, the relations of  $\omega-\omega'$ , $\omega-z$ and $\omega'-z$ are list  in Figure \ref{wwzp}. The curves show  the deviations from the $\Lambda$CDM model are
tiny with small $f$.
Specifically, the knowledge of $f$ is suffice to know the present value of EoS parameter where
$1+\omega_{0}=+4f/3(1+f)$ and $\omega_{0}'=-16f/3(1+f)^{2}$.
 If  $f$ is at the order of $10^{-5}$, it is not surprising that the EoS parameter
is very close to $-1$ and the derivative of  the EoS parameter is tiny.

\begin{figure} \centering
 {\includegraphics[width=2.0in]{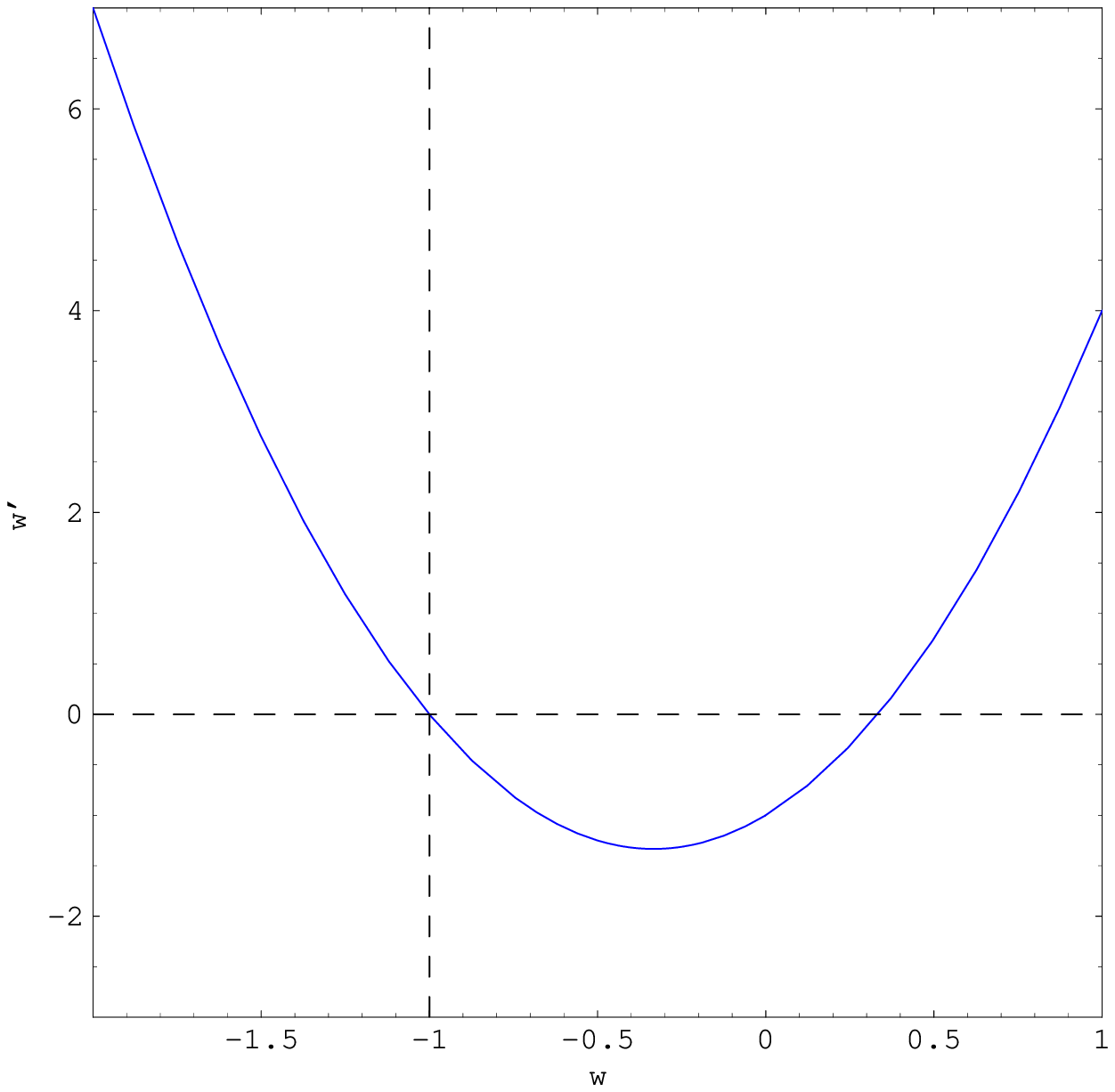}}\quad
{\includegraphics[width=2.0in]{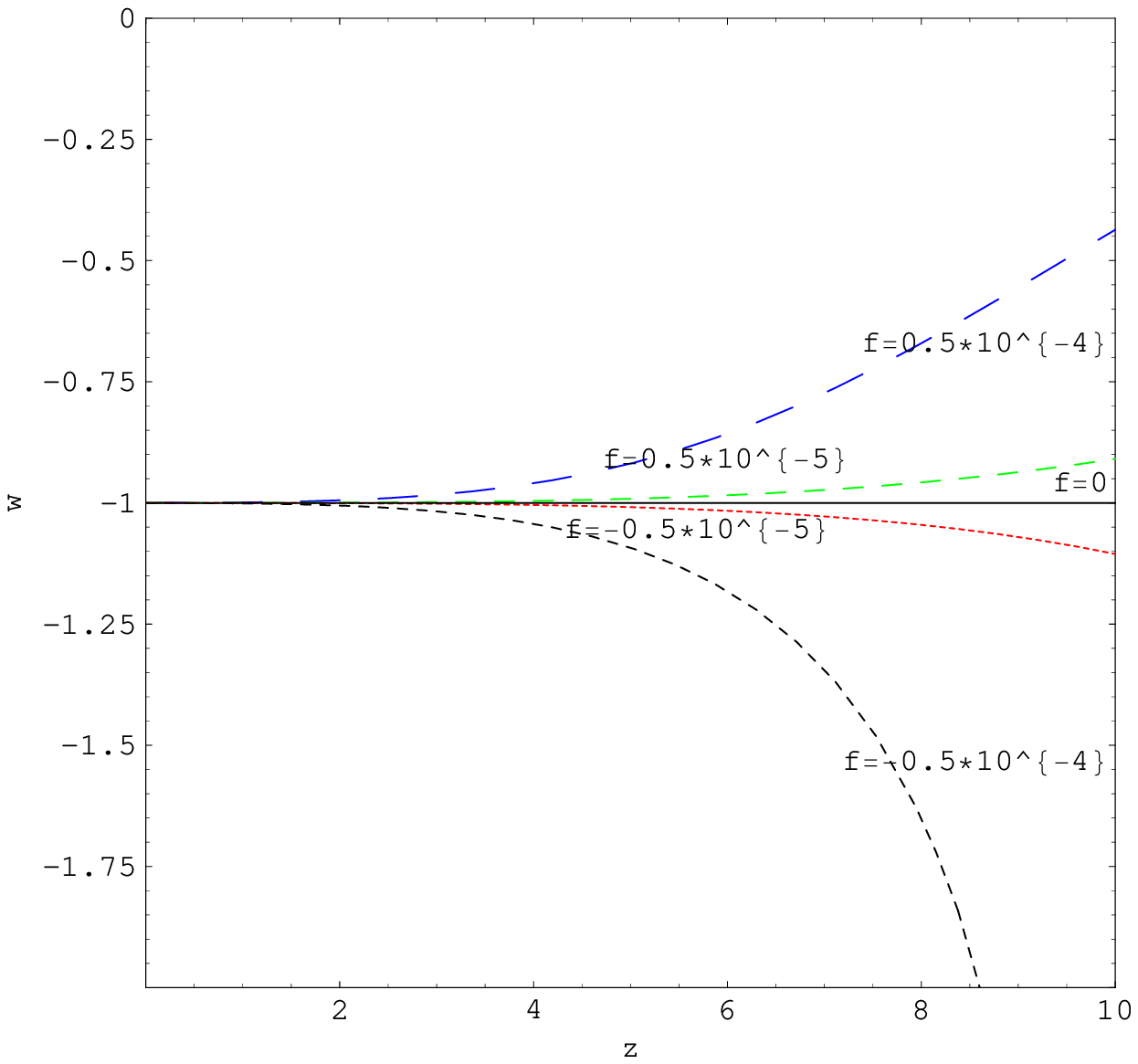}}\quad
 {\includegraphics[width=2.0in]{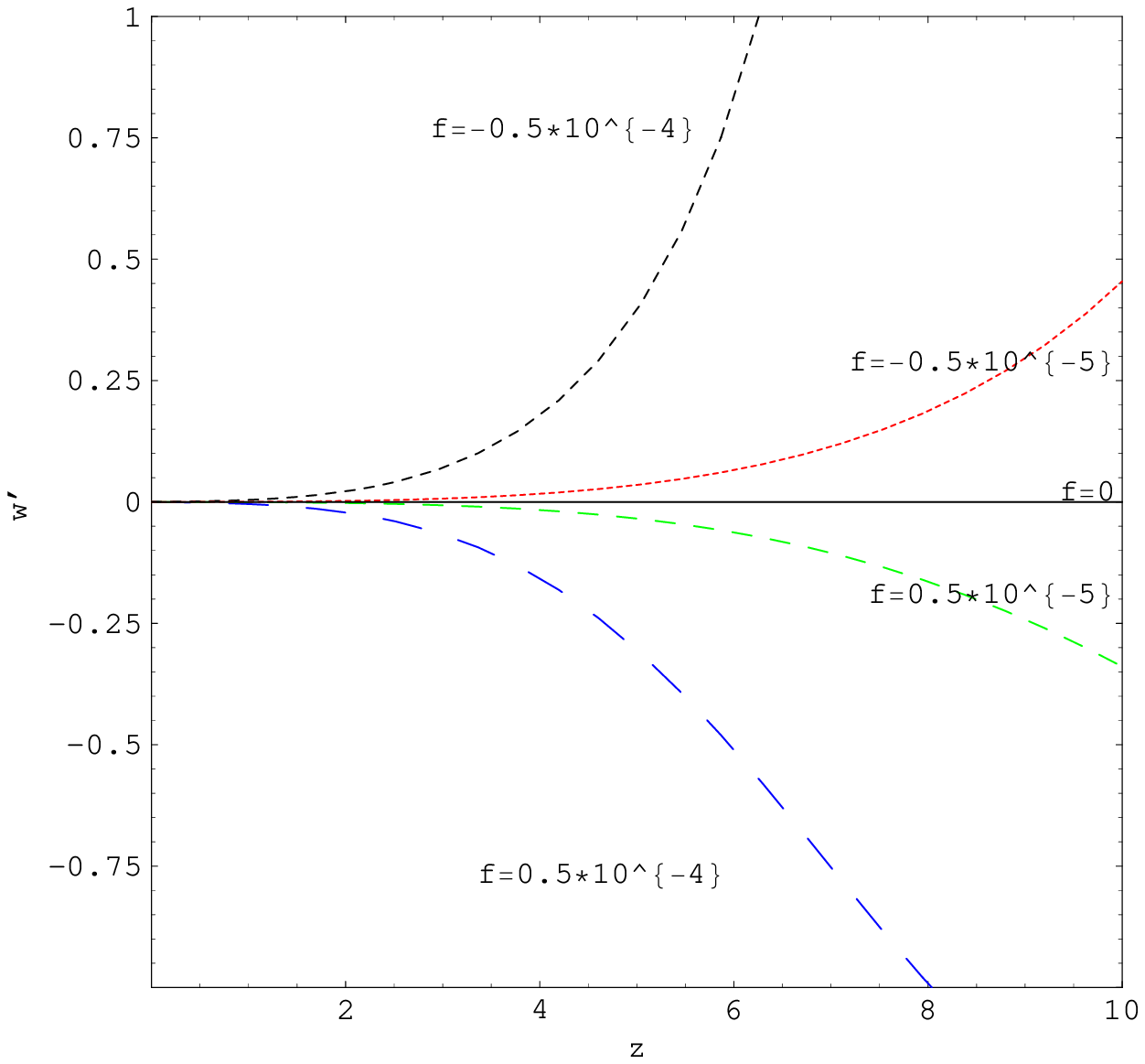}}\quad

\caption{The left, middle and right panels are the relation of $\omega-\omega'$,  $\omega-z$ and $\omega'-z$  separately. }\label{wwzp}
\end{figure}

\section{The Phase-Space Analysis}\label{3}
To do phase-space analysis in our model, three
dimensionless parameters are defined firstly,
\begin{eqnarray}
&&u=\sqrt{\frac{H_{0}^{2}\Omega_{m0}a^{-3}}{H^{2}}},\\
&&v=\sqrt{\frac{H_{0}^{2}(\Omega_{r0}+\Omega_{dr0})a^{-4}}{H^{2}}},\\
&&w=\sqrt{\frac{H_{0}^{2}\Omega_{k0}a^{-2}}{H^{2}}}.
\end{eqnarray}
 Based the Friedmann equation and the conserved ones
  \footnote{The conserved equation for each component is $\dot{\rho}_{i}+3H\rho_{i}(1+\omega_{i})=0$ where $\omega_{r}=1/3$, $\omega_{dr}=1/3$, $\omega_{m}=0$ and $\omega_{d}=-1$.},
  the evolutions of  $u,v,w$ are
\begin{eqnarray}
\label{x}
&&u'=u(-\frac{3}{2}+\frac{3}{2}u^{2}+2v^{2}+w^{2}),\\
\label{v}
&&v'=v(-2+\frac{3}{2}u^{2}+2v^{2}+w^{2}),\\
\label{w}
&&w'=w(-1+\frac{3}{2}u^{2}+2v^{2}+w^{2}).
\end{eqnarray}
When $u'$, $v'$ and $w'$ are all equal to $0$, the corresponding value of $u,v, w$ gives a critical point.
Four points are  list  in the Table \ref{tab1}. And, we could put a small perturbation near the critical points' neighbor. Then,  the perturbation equations are gotten.  If  the real parts of the eigenvalues of the perturbation equations  are all  positive, the
corresponding critical point is an unstable fixed point. In contrast,
the negative real parts of the eigenvalues  denotes a stable point.  Specially if the real parts of the
eigenvalues  are mixed with the negative one and the positive one, the
corresponding critical point is an unstable saddle point \cite{dynamics,Wainwright,Coley}.

 Generally speaking, the model with dark radiation   could go through
 the unstable radiation dominated phase ($R$), the unstable matter dominated phase ($M$),
 the stable dark energy dominated phase ($D$) and the unstable
 curvature dominated phase ($K$) \footnote{The unstable curvature dominated phase often represents some very early physics. Here, we ignore this phase on the discussions of the history of our universe.}.
The dark energy dominated phase is stable which means the universe
will  be dominated by the cosmological constant in the future. Before that, our universe is supposed to  go through
these unstable phases.  This process is corresponding to the history of our universe:
the radiation dominated phase at first, then matter dominated phase and finally
the dark energy dominated phase.

\begin{table*}[t]
\begin{center}
\begin{tabular}{|llllll|}
\hline Phase& Physical Meaning& $(u,v,w)$ &Existing  & $(\lambda_1,\lambda_2,\lambda_3)$ & Stability  \\
\hline $M$& Matter Dominated & $(1,0,0)$ &Always  &$(3,-2,-1)$ & Unstable \\
\hline $R$&  Radiation Dominated & $(0,\pm1,0)$ &Always &$(\frac{3}{2},4,-1)$ & Unstable \\
\hline $K$&  Curvature Dominated & $(0,\pm1,0)$ &Always &$(-\frac{3}{2},-2,2)$ & Unstable \\
\hline $D$&  Dark Energy Dominated & $(0,0,0)$ &Always &$(-\frac{3}{2},-2,-1)$ & Stable \\
\hline
\end{tabular}
\end{center}
\caption[crit]{\label{tab1}We list the properties of the critical points: the physical meaning of  the phases, the value of the phases, the existing condition, the eigenvalues of the points and the stability of the phases. }
\end{table*}

\section{The Data and The  Method Analysis}\label{sec4}

Once treating dark radiation as  signal of dark energy,
   the observational testing method used in the dark energy model could be applied  to test the dark radiation.
   In this section,  the data and the analytical methods will be presented  separately. 

\subsection{The Data Analysis}
\subsubsection{The $SNLS$ data}
SNe Ia (supernovae) is used in the standard distance method which measures the expansion of our universe.
For the $SNLS$ data, Ref. \cite{Conley:2011ku} gives
the apparent $B$ magnitude $m_B$,  and the covariance matrix for
$\Delta m \equiv m_B-m_{\rm mod}$,  with
\begin{eqnarray}
m_{\rm mod}=5 \log_{10} D_L(z|\mbox{ s})
- \alpha (s-1) +\beta {\cal C}_{SN} + {\cal M},
\end{eqnarray}
where $ D_L(z|\mbox{ s})$ is the luminosity distance
multiplied by $H_0$
for a given set of cosmological parameters $ s$ \footnote{
$s$ is the stretch measure of the SNe Ia light curve shape.},
${\cal C}_{SN}$ is the color measure for supernovae and
${\cal M}$ is a nuisance parameter representing some combination
of the absolute magnitude of a fiducial SNe Ia.
The time dilation part of the observed luminosity distance depends
on the total redshift $z_{\rm hel}$ \cite{Hui:2005nm}
\begin{eqnarray}
D_L(z|\mbox{\bf s})\equiv c^{-1}H_0 (1+z_{\rm hel}) r(z|\mbox{\bf s}),
\end{eqnarray}
where $c$ is the color index, $z$ and $z_{\rm hel}$ are the CMB rest frame and heliocentric redshifts
of the supernovae. The correlated errors is
\begin{eqnarray}
\chi^2_{SN}=\Delta  m^T \cdot \mbox{\ C}_{SN}^{-1} \cdot \Delta m,
\end{eqnarray}
where
$\mbox{ C}_{SN}$ is the $N\times N$ covariance matrix of the SNe Ia where $N$ is the number of the components.
The nuisance parameter
$H_0$ is marginalized over by evaluating $\chi^2_{SN}$.

\subsubsection{The CMB Data}
 The CMB
data is implemented to add distance measurements at higher redshift ($z>10$).
We  use the derived quantities of the $WMAP9$ and $PLANCK$ measurements  \cite{Wang:2011sb,Wang:2013mha}: the shift parameter
$R(z^{*})$, the acoustic scale $l_A(z^{*})$ at the decoupling redshift and the base parameter $\omega_b$ whose definition is $\Omega_{b}h^{2}$ where $\Omega_b$ is the fractional energy densities for baryon. 
The  $\chi^2$ of CMB data is
\begin{equation}
\label{cmbchi} \chi^2_{CMB}(\mathbf{p},\Omega_b h^2, h)=\sum_{i,j=1}^{3}\Delta x_i C_{CMB}^{-1}(x_i,x_j)\Delta x_j,
\end{equation}
where the three parameters are $x_i=(R(z^{*}), l_A(z^{*}), \omega_b)$, $\Delta
x_i=x_i-x_i^{obs}$ and $C_{CMB}(x_i,x_j)$ is the covariance matrix
for the three parameters \cite{Komatsu:2010fb,Wang:2011sb,Wang:2013mha}. The shift parameter $R$ is
expressed as $R(z^{*})=\left[\sqrt{\Omega_{m}}\rm
sinn(\sqrt{|\Omega_{k}|}\int_0^{z^{*}}dz/E(z))\right]/\sqrt{|\Omega_{k}|}=1.710\pm
0.019$.
 The acoustic scale  is
$l_{A}(z^{*})=\pi
d_{L}(z^{*})/(1+z^{*})r_{s}(z^{*})=302.1\pm0.86 $.
And the decoupling redshift $z^{*}$ is fitted by Ref.\cite{Hu:1995en} with
$ z^{*}=1048[1+0.00124(\Omega_{b} h^{2})^{-0.738}][1+g_{1}(\Omega_{m} h^{2})^{g_{2}}] =1090.04\pm 0.93$,
where
$g_{1}=0.0783(\Omega_{b} h^{2})^{-0.238}[1+39.5(\Omega_{b}
h^{2})^{0.763}]$ and $ g_{2}=0.560/[1+21.1(\Omega_{b} h^{2})^{1.81}]$.

\subsubsection{The $BAO$, $H(z)$ and $BBN$ Data}
To produce tightest cosmological constraint, we try to use
 other cosmological probes as well.

The $BAO$ data  are used as standard rule. Due
to the sound waves in the plasma of the early universe,
 the
 wavelength of $BAO$
is related to the co-moving sound horizon at the baryon drag epoch which is
$ d_{z}= r_{s}(z_{d})/D_{V}(z)$,
 $D_V$ is the effective distance with
$D_V(z)=\left[z d_L^2(z)/H(z)(1+z)^2\right]^{1/3}$,
$z_d$ is the drag redshift defined in \cite{Eisenstein:1997ik},
and  $
r_s(z)=\int_z^\infty c_s(x)dx/E(x)$ is the co-moving sound horizon \footnote{ The sound speed is $c_s(z)=1/\sqrt{3[1+\bar{R_b}/(1+z)}]$ where
$\bar{R_b}=3\Omega_b h^2/(4\times2.469\times10^{-5})$.}.
For the BAO data, we use the measurements from the 6dFGS (hereafter Bao1) \cite{Beutler:2011hx}),
 the distribution of galaxies ( hereafter Bao2) \cite{Percival:2009xn} and the WiggleZ dark
energy survey (hereafter Bao3) \cite{Blake:2011en}. The 6dFGS (Bao6dF) gives
\begin{eqnarray}
\chi^{2}_{Bao1}=\frac{(d_{0.106}-0.336)^2}{0.015^2}.
\end{eqnarray}
And,  the distribution of galaxies (Bao2)  measured the distance ratio
at two redshifts $z=0.2$ and $z=0.35$, whose $\chi^{2}$ is
 \begin{eqnarray}
\chi^{2}_{Bao2}=\sum_{i,j=1}^{2}\Delta d_i
C_{dz}^{-1}(d_i,d_j)\Delta d_j
\end{eqnarray}
where $d_i=(d_{z=0.2},d_{z=0.35})$, $\Delta d_i=d_i-d_i^{obs}$ and
the covariance matrix $C_{dz}(d_i,d_j)$ for $d_z$ at $z=(0.2,0.35)$
is taken from Eq. 5 in Ref.\cite{Percival:2009xn}.
Furthermore, the WiggleZ dark energy survey measured the acoustic parameter
$A(z)=D_V(z)\sqrt{\Omega_m H_0^2}/z$
at three redshifts $z=0.44$, $z=0.6$ and $z=0.73$, and the results and their
 covariance matrix are listed
in Table 2 and Table 3 of Ref.\cite{Blake:2011en}. The $\chi^{2}$ is
\begin{eqnarray}
\chi^{2}_{Bao3}=\sum_{i,j=1}^{3}\Delta A_i C_{A}^{-1}(A_i,A_j)\Delta A_j,
\end{eqnarray}
where $A_i=[A(0.44),A(0.6),A(0.73)]$, $\Delta A_i=A(z_i)-A(z_i)^{obs}$. Then, the total $\chi^{2}$ for $BAO$ is
\begin{eqnarray}
\label{baochi2}
\chi^2_{Bao}(\mathbf{p},\Omega_b h^2, h)=\chi^2_{Bao1}+\chi^2_{Bao2}+\chi^2_{Bao3}.
\end{eqnarray}

To alleviate the double integration  of the EoS
parameter $\omega(z)$,
 we also apply the measurements of the Hubble parameter $H(z)$
which could better constrain $w(z)$ at high redshift.
We use the $H(z)$ data at 11 different redshifts obtained from the differential ages of
red-envelope galaxies in Ref.\cite{Simon:2004tf}, and three more Hubble parameter data
$H(z=0.24)=76.69\pm2.32$, $H(z=0.34)=83.8\pm2.96$ and
$H(z=0.43)=86.45\pm3.27$ determined by Ref.\cite{Riess:2009pu}. Then, the $\chi^2$ of Hubble parameter  data is
\begin{equation}
\label{hzchi}
\chi^2_H(\mathbf{p}, h)=\sum_{i=1}^{14}\frac{[H(z_i)-H_{obs}(z_i)]^2}{\sigma_{hi}^2},
\end{equation}
where $\sigma_{hi}$ is the $1\sigma$ uncertainty in the $H(z)$
data.

Furthermore,  the constraint data from $BBN$  is added for this dark radiation model.
The  $\chi^{2}$ of the  Big Bang Nucleosynthesis ($BBN$) data   \cite{Serra:2009yp,Burles:2000zk,Steigman:2005uz})  is
\begin{equation}
\chi_{bbn}^{2}=\frac{\left(\Omega_{b0}h^{2}-0.022\right)^{2}}{0.002^{2}},
\end{equation}
where $\Omega_{b0}=0.02253h^{2}$ is the present value of dimensionless energy density for baryon.

\subsubsection{Data Discussion}

 To use the data properly,  the $SNLS$ data will be used individually at first and be denoted as ``Data  \Rmnum{1} ".
 To utilize the  $WMAP9$ and $PLANCK$ data  seperately, we treat $WMAP9+BAO+H(z)+BBN$ and $PLANCK+BAO+H(z)+BBN$ as 
 ``  Data \Rmnum{2}" and ``  Data \Rmnum{3}".
 If all the three data are consistent, then we combine them as ``Data \Rmnum{4}: $SNLS+WMAP9+PLANCK+BAO+H(z)+BBN$".

\subsection{The Analysis Method}

 Monte Carlo Markov Chain (MCMC) method is used  to compute
the likelihood for the parameters  in the
model.
By using the Metropolis-Hastings algorithm, the MCMC method
randomly chooses values for the parameters, evaluates $\chi^2$
and determines whether to accept or reject the set of parameters.

\subsubsection{The $AIC$ Principle}

After combining different data,   the total $\chi^{2}$ could be gotten by adding all the observation's $\chi^{2}$ together.
The model parameters  are determined by minimizing $\chi^{2}$.
To value the goodness of  fitting,  the $AIC$ (Akaike Information Criterion) principle will be used  which is very popular in Mathematics and Physics \cite{Akaike,Liddle:2004nh},
\be
AIC= \chi_{min}^{2}+2n,
\ee
where $n$ is the number of parameters and $\chi_{min}^{2}$ is the minimum value of $\chi^{2}$.
The smaller the $AIC$ value is, the better the constraint is. If the $\chi^2$  difference  between two models is in a range of  $0<\Delta(AIC)<2$, the constraints of the two models are considered to be equivalent.

\subsubsection{The $Om$ Diagnostic}
 The
 $Om$ diagnostic \cite{Sahni:2008xx,Shafieloo:2012rs} is proposed to distinguish dynamical dark energy
 from the cosmological constant 
 both with and without the matter density. In another saying, it is on the basis of observations of the expansion
 history.
The $Om$ diagnostic   could be characterized by
 \begin{eqnarray}
Om(x)=\frac{H^{2}(x)/H_{0}^{2}-1}{x^{3}-1}, \,\,\,\,\,x=1+z.
 \end{eqnarray}

\begin{table*}[t]
\begin{center}
\begin{tabular}{lllll}
& Flat  $\Lambda CDM$  &Curved  $\Lambda CDM$ & Flat  DR Model & Curved DR Model\\
\hline \multicolumn{5}{l}{Data \Rmnum{1}: $SNLS$}\\
$\Omega_{m0}$ & $0.226^{+0.039+0.070}_{-0.037-0.070}$&$0.174^{+0.138+0.222}_{-0.154-0.169}$&$0.297^{+0.306+0.383}_{-0.292-0.292}$&$0.145^{+0.832+0.835}_{-0.138-0.140}$  \\
$f$&$-$&$-$&$-0.68^{+0.51+0.90}_{-0.12-0.93}$&$0.22^{0.15+0.19}_{-0.74-0.90}$  \\
 $\Omega_{k0}$&$-$&$0.150^{+0.41+0.59}_{-0.36-0.56}$&$-$&$0.17^{+0.52+0.62}_{-0.82-0.88}$ \\
 $N_{eff} (10^{4})$&$-$&$-$&$-0.36_{-1.34-1.65}^{\,\,+1.61+1.91}$&$0.12_{-2.78-2.78}^{\,\,+1.06+1.54}$\\
 $\chi^{2}$ &$420.10$&$419.77$&$418.60$ & $419.78$ \\
 $AIC$ &$422.10$&$423.77$&$422.60$ & $425.78$ \\

  \hline \multicolumn{5}{l}{Data \Rmnum{2}: $WMAP9+BAO+H(z)+BBN$} \\
$\Omega_{m0}$  &$0.280_{-0.025-0.036}^{+0.029+0.044}$&$0.280_{-0.028-0.039}^{+0.034+0.050}$ &$0.286_{-0.033-0.045}^{+0.040+0.059}$& $0.287_{-0.038-0.051}^{+0.049+0.070}$  \\
$f(10^{-5})$ &$-$&$-$ &$0.37_{-1.04 -1.49}^{+1.13+1.63}$& $0.45_{-1.60-2.14}^{+1.83+2.63}$   \\
 $\Omega_{k0} (10^{-2})$ &$-$&$0.18_{-0.93-1.31}^{+0.99+1.40}$ &$-$&$-0.09_{-1.55-2.19}^{+1.46+1.97}$  \\
  $H_{0}$ & $70.19_{-2.29-3.40}^{+2.21+3.31}$ &$70.43_{-2.88-4.11}^{+2.86+4.02}$ &$70.73_{-3.05-4.27}^{+3.04+4.35}$& $70.74_{-3.43-4.67}^{+3.38+4.62}$\\
 $N_{eff} $ &$-$&$-$ &$3.28_{ -0.65-0.92 }^{+0.72+1.04}$& $3.33_{-0.10-1.32}^{+1.16+1.63}$\\
 $\chi^{2}$  &$10.17$&$9.95$ &$9.55$& $9.54$ \\
 $AIC$ &$12.17$&$13.95$ &$13.55$& $15.54 $ \\ 
 
 \hline \multicolumn{5}{l}{Data  \Rmnum{3}: $PLANCK+BAO+H(z)+BBN$} \\
$\Omega_{m0}$  &$0.290_{-0.020-0.030}^{+0.024+0.036}$&$0.286_{-0.024-0.333}^{+0.290+0.417}$ &$0.296_{-0.026-0.037}^{+0.030+0.044}$& $0.292_{-0.039-0.050}^{+0.047+0.068}$  \\
$f(10^{-5})$ &$-$&$-$ &$0.44_{-0.98-1.42}^{+1.00+1.43}$&  $0.29_{-1.48-2.01}^{+1.78+2.55}$ \\
 $\Omega_{k0} (10^{-2})$&$-$&$0.30_{-0.71-1.02}^{+0.69+0.97}$ &$$& $0.17_{-1.29-1.81}^{+1.05+1.38}$\\
 $H_{0}$ & $69.46_{-1.81-2.67}^{+1.75+2.60}$ &$70.35_{-2.90-4.11}^{+2.76+3.94}$ &$70.31_{-2.80-4.01}^{+2.95+4.15}$& $70.46_{-3.29-4.54}^{+3.38+4.56}$\\
 $N_{eff} $ &$-$&$-$ &$3.32_{-0.60 -0.84}^{+0.65+0.93}$& $3.23_{ -0.92-1.24}^{+1.09+1.54}$\\
 $\chi^{2}$  &$10.73$&$9.84$ &$9.71$& $9.62$ \\
 $AIC$ &$12.73$&$13.84$ &$13.71$& $15.62$ \\

 \hline \multicolumn{5}{l}{Data \Rmnum{4}: $SNLS+WMAP9+PLANCK+BAO+H(z)+BBN$} \\
 $\Omega_{m0}$  &$-$&$0.281_{-0.027-0.035}^{+0.029+0.041}$ &$0.286_{-0.025-0.032}^{+0.031+0.042}$& $0.280_{-0.036-0.045}^{+0.049+0.063}$  \\
$f(10^{-5})$ &$-$&$-$ &$0.31_{-1.11-1.46}^{+1.10+1.50}$&  $0.08_{-1.53-1.86}^{+1.82+2.43}$\\
 $\Omega_{k0} (10^{-2})$ &$-$&$0.26_{-0.77-1.07}^{+0.75+0.95}$ &$-$& $0.32_{-1.40-1.81}^{+0.95+1.24}$   \\
   $H_{0}$ & $-$ &$70.65_{-3.10-4.09}^{+3.39+4.25}$ &$70.78_{-3.42-4.49}^{+3.24+ 4.17}$& $71.11_{-3.90-5.01}^{-3.45-4.45}$\\
 $N_{eff} $ &$-$&$-$ &$3.25_{-0.68-0.88}^{+0.74+1.00}$& $3.09_{-0.97-1.18}^{+1.17+1.53}$\\
 $\chi^{2}$  &$-$&$433.62$ &$433.86$& $433.66$ \\
 $AIC$ &$-$&$437.62$ &$437.86$& $439.66$ \\
\hline
\end{tabular}
\end{center}
\caption[crit]{\label{tab} The maximum likelihood
 values with $1\sigma$ and $2\sigma$ confidence ranges  are presented  for the flat  $\Lambda CDM$ model, the curved $\Lambda CDM$ model,
 the flat $\Lambda CDM$ model with dark radiation (Flat DR Model) and the curved $\Lambda CDM$ model with dark radiation (Flat DR Model).}
\end{table*}

\section{The Fitting Results}\label{sec5}

\subsection{The Flat and Curved $\Lambda$CDM Model}
To  a certain  model, different data may  give out very different constraining  results  which is called tension.
One reason of  tension is the system error of different  data.
 The other reason could be traced to the model which may not represent the true physics.

For flat $\Lambda CDM$ model, Table \ref{tab}  shows the $1\sigma$ upper bound of $\Omega_{m0}$
 given by the SNLS data  is  $0.265$ which is incompatible with the $1\sigma$ lower bound of $\Omega_{m0}$
  given by Data \Rmnum{3} where $\Omega_{m0}= 0.270$. In another saying,  the constraining parameter range from the two Data sets   are not  overlapped at $1\sigma$ level.  For Data \Rmnum{2}, this situation are slightly better  where the lower bound of $\Omega_{m0}$ is $0.252$. Anyway, that is just slightly  overlapped with the SNLS data.   Tension exists between the $SNLS$ data and the other data (including $WMAP9$, $PLANCK$, $H(z)$ and $BBN$). And,
 we could not combine   all the data together for flat $\Lambda CDM$ model.

Fortunately, after adding the curvature, 
   Table \ref{tab} shows all the data are consistent.  The $1\sigma$ range values of $\Omega_{m0}$
    are overlapped and the
   tensions between  the $SNLS$ data and the other data are disappeared.  Thus, it is reasonable to combine Data   \Rmnum{1},  \Rmnum{2} and  \Rmnum{3} to get tighter constraints for  the curved $\Lambda$CDM model.
Comparing the flat and curved $\Lambda CDM$ model,  $\Delta (AIC)=1.67<2$; so  the constants of the two models are considered to be equivalent.  This tension resolution  hints  the system error of the data may not the reason. Refs.\cite{Clarkson:2007bc,Okouma:2012dy} show that the assumption of a flat universe induces critically large errors in reconstructing the dark energy equation of state  even if the true cosmic curvature is very small.
As the dark radiation  is also a small component, in the following,  we will try to answer the question  that whether the dark
radiation part could alleviated the tension problem or not  based on observations\footnote{
The effect that extra radiation can smash off the data tension are reported for other observational data,  e.g. \cite{Godlowski:2006vf,Mangano:2006ur,Bashinsky:2003tk,Archidiacono:2011gq}.}.

 \subsection{ The Flat and Curved $\Lambda$CDM Model with Dark Radiation }
 \begin{figure} \centering
  {\includegraphics[width=2.4in]{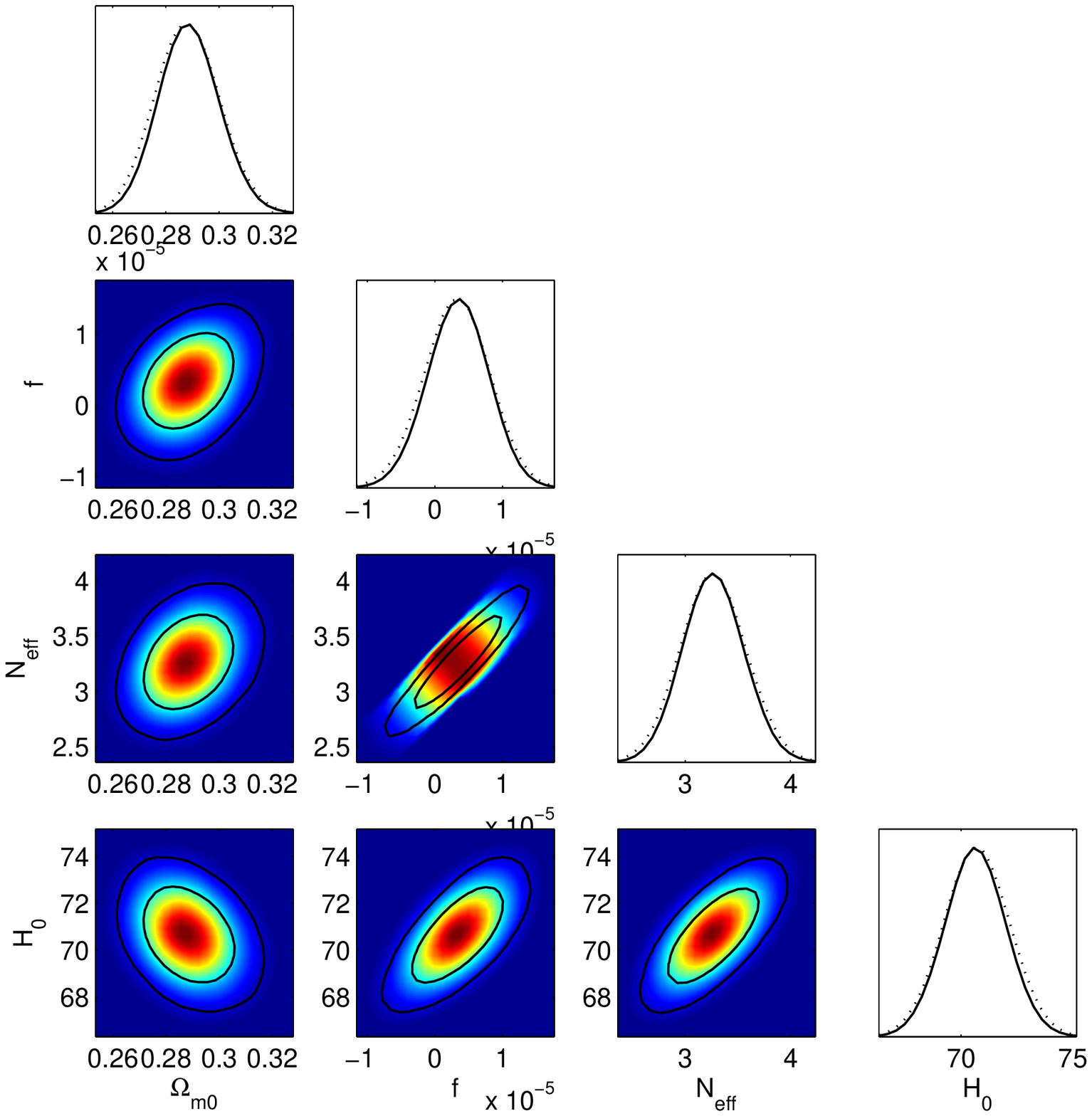}\label{nokt}}\quad
   {\includegraphics[width=2.4in]{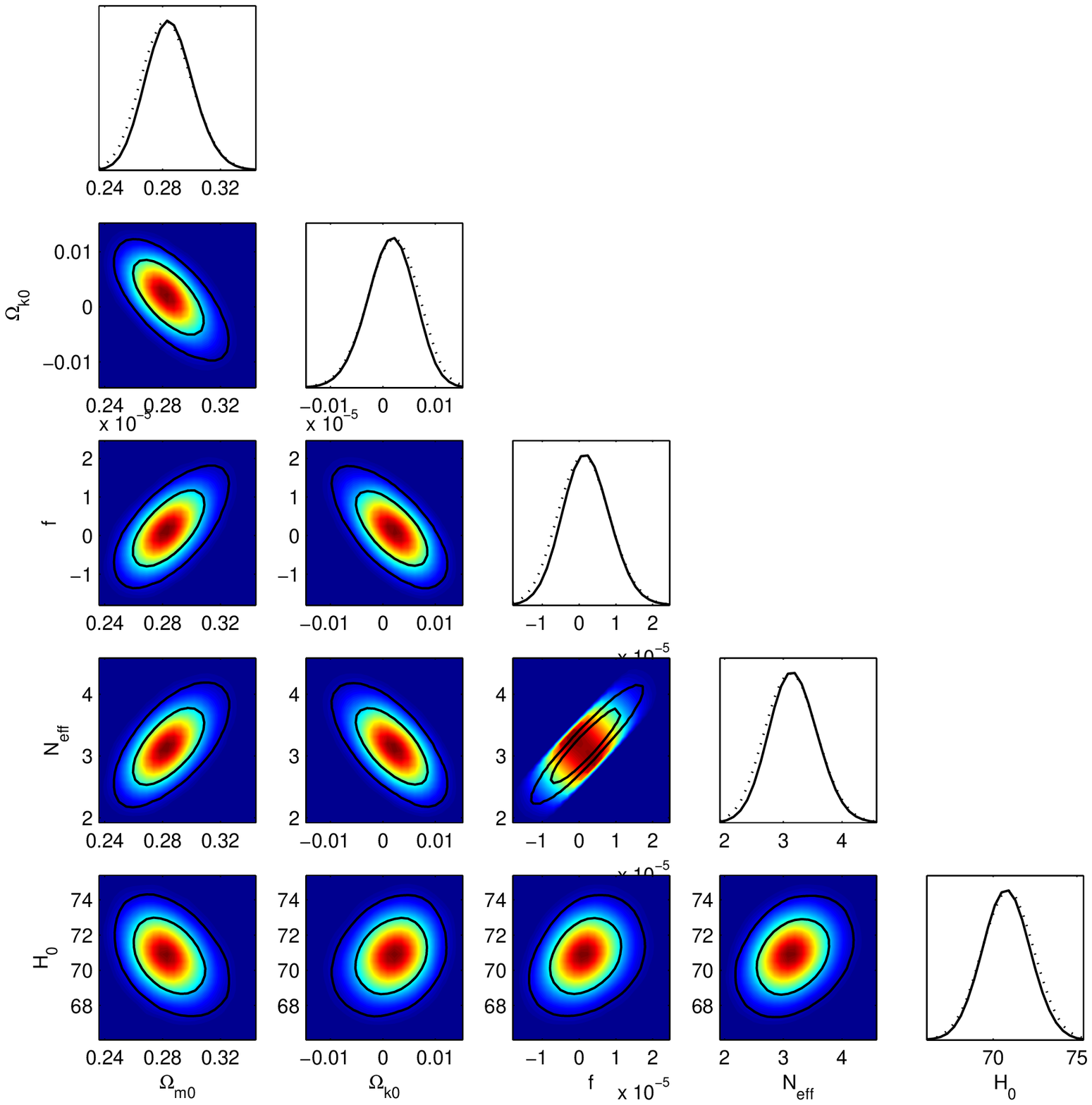}\label{kt}}\quad
\caption{ The constraining results of  the flat and curved dark radiation models are presented in the left and right panels separately? The results  are given by the $SNLS+WMAP9+PLANCK+BAO+H(z)+BBN$ data (Data \Rmnum{4}). } \label{tri}
\end{figure}

\begin{figure} \centering
  {\includegraphics[width=2.4in]{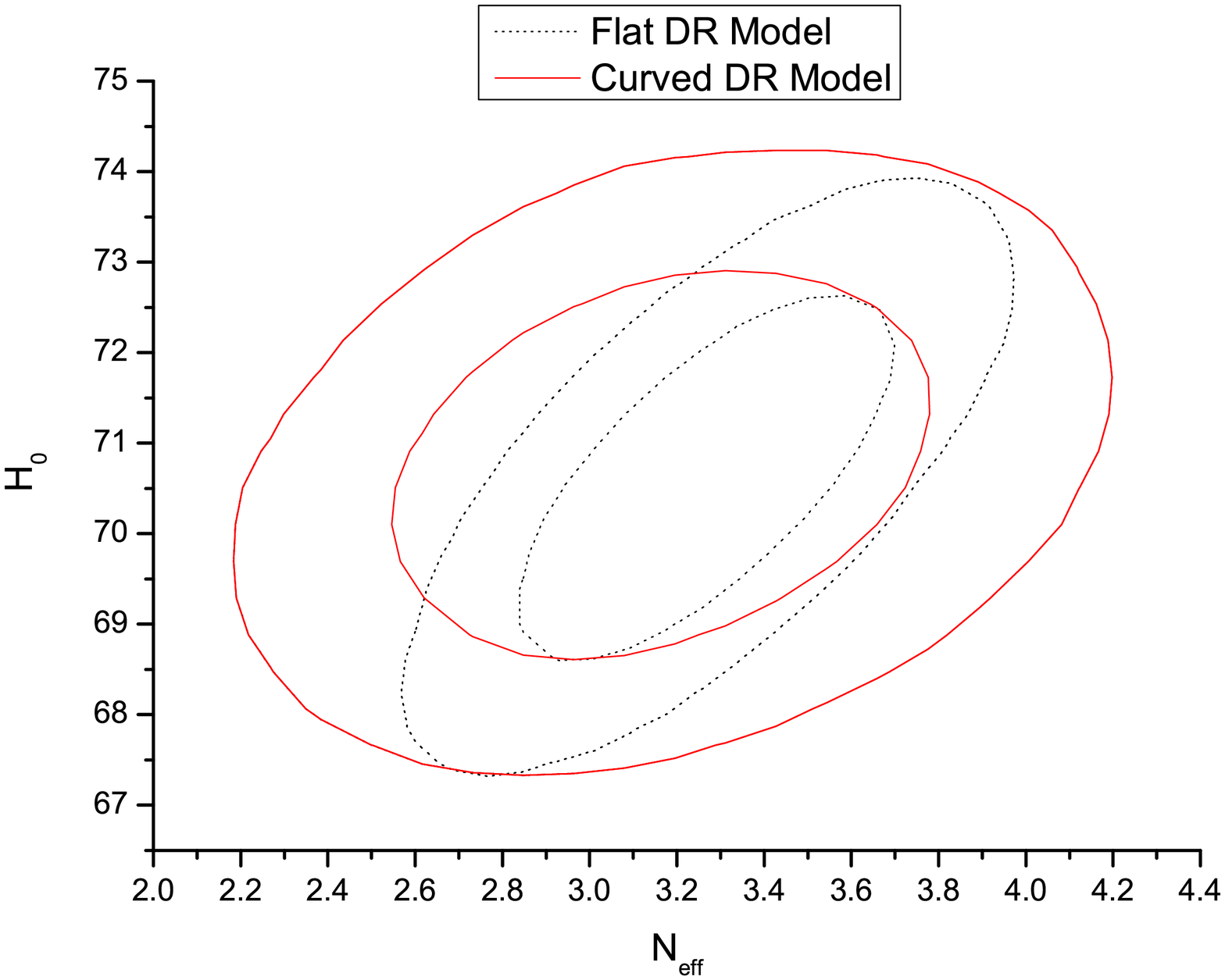}}\quad
\caption{ The  contours for  $H_{0}$ and $N_{eff}$ are given out.}\label{HN}
\end{figure}

 After adding the dark radiation to the $\Lambda$CDM model,
 Table \ref{tab} shows
 $SNLS$ data gives a loose parameter range.
 As we expected, this tension problem is disappeared. Therefore, it is reasonable to use the combined $SNLS+WMAP9+PLANCK+BAO+H(z)+BBN$ data. It  gives out the tightest constraints.
Then, what results could we get if we add  both the curvature and the dark radiation to the $\Lambda$CDM model? As shown in  Figure \ref{tri}, the parameter ranges of the curved one are slightly enlarged compared to the flat one.
 
 Generally speaking, the $SNLS $ data gives very poor constraints on the model parameter compared to other data.    Data \Rmnum{2}, \Rmnum{3} and \Rmnum{4} present 
$\Omega_{k0}\sim10^{-2}$ and $f\sim 10^{-5}$ which denote the price we paid for the disappeared tension is reasonable.  The AIC analysis also shows the constraints on both the flat and curved ones are equal because the $\Delta (AIC)$ is less than $2$. Anyway, the $PLANCK+BAO+H(z)+BBN$
 data gives a tighter constraint than the $WMAP9+BAO+H(z)+BBN$ data.

\subsubsection{The Dark Radiation}

 Again, as Table \ref{tab} hints,  the $SNLS$ data is not
sensitive to the effective neutrino number.  In contrast, the  constraints from other data are at much smaller orders.  For concise, we only discuss the
tightest constraints from Data \Rmnum{4}. The combined data favor a positive $f$ which  denotes the
 new produced dark radiation. Based on our definition,
 the data gives out
 $N_{eff}=3.25_{-0.68-0.88}^{+0.74+1.00}$ in flat dark radiation model and
$N_{eff}= 3.09_{-0.97-1.18}^{+1.17+1.53}$ in the curved one. Dark radiation makes  the best fit of $N_{eff}$  slightly larger than $3.04$.  We compare  the flat  and  curved cases by drawing  the contours of $H_{0}$ and $N_{eff}$ in Figure \ref{HN} where the ranges of $H_{0}$  are nearly the same, but   the  range of $N_{eff}$ is larger in the curved case.

\subsubsection{The   $Om$ Diagnostic}

The $Om$ diagnostic is used to distinguish the dark radiation effect. For our model, 
 \begin{eqnarray}
Om(x)=\Omega_{m0}+\frac{(\Omega_{r0}+\Omega_{dr0})(x^{2}+1)(x+1)}{x^{2}+x+1}+\frac{\Omega_{k0}(1+x)}{x^{2}+x+1}.
 \end{eqnarray}
 Generally, the effect of today's dark radiation makes $\delta Om_{dr0}<4\Omega_{dr0}/3$. Meanwhile, the effect of today's curvature makes $\delta Om_{k0}<\Omega_{dk0}/2$. As $f$ (or $\Omega_{dr0}$) are relative small,  $\delta Om_{dr0}$ is  smaller than $\delta Om_{k0}$.   Robustly,
 $\Omega_{dr0}$ ($\sim10^{-5}$) is four orders smaller than $\Omega_{m0}$ ($10^{-1}$) and three order smaller than $\Omega_{k0}$ ($10^{-2}$). The relation of $Om-z$
are drawn out  in Figure \ref{Om} for the
flat and curved dark radiation cases. In the flat one,
 the best fitting value of $Om$  is nearly constant, so does its $1\sigma$ and $2\sigma$ ranges.
 But the behavior of $Om$ in the curved $\Lambda CDM$ model and the curved dark radiation model show dynamical signals.
 Considering  the  flat $\Lambda CDM$ model which gives a constant $Om$ \cite{Sahni:2008xx} as well,
  the flat dark
  radiation  model could not  be distinguished from the flat $\Lambda CDM$ model while the curved dark
  radiation  model can.

\begin{figure} \centering
{\includegraphics[width=2.0in]{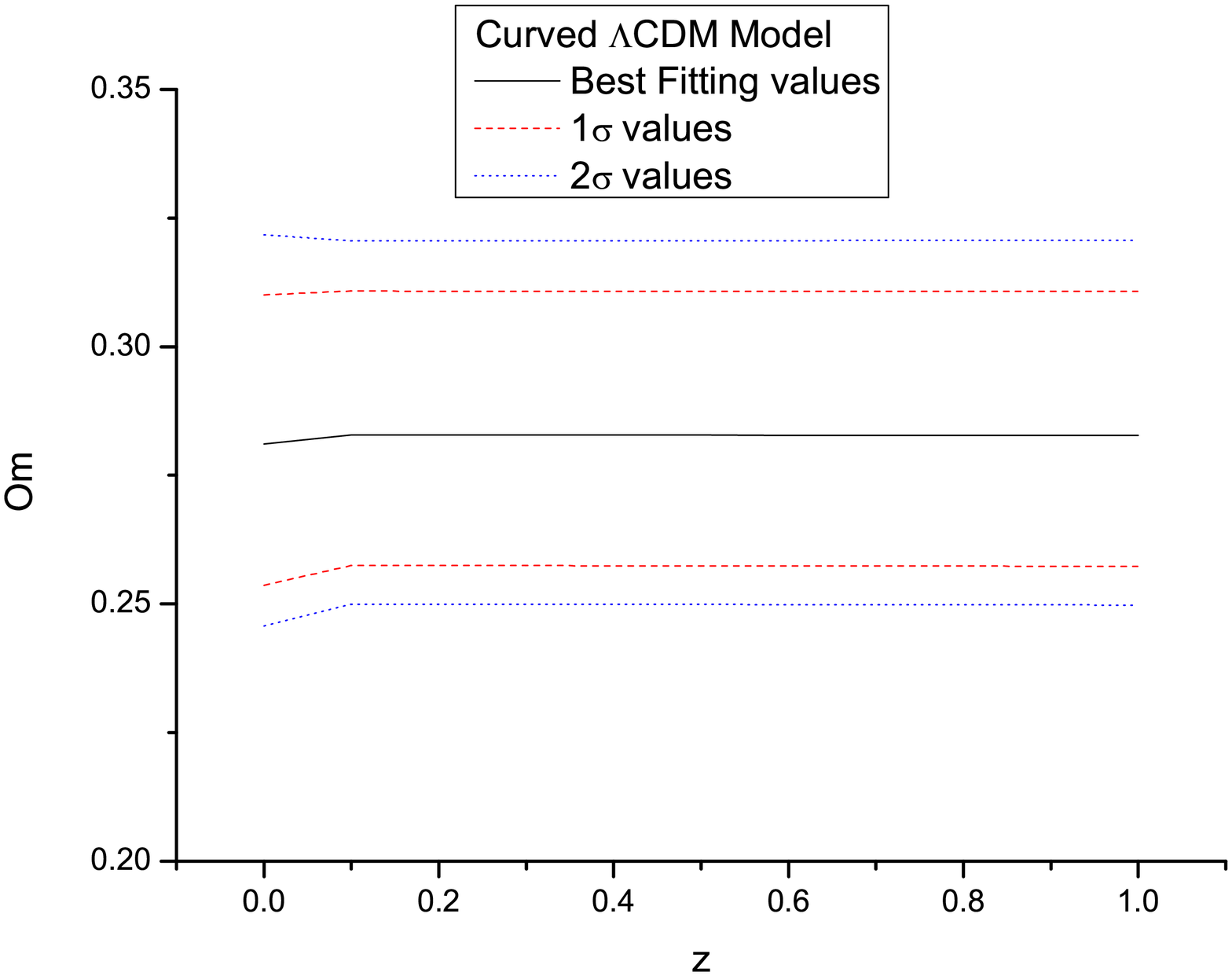}\label{omflat}}\quad
 {\includegraphics[width=2.0in]{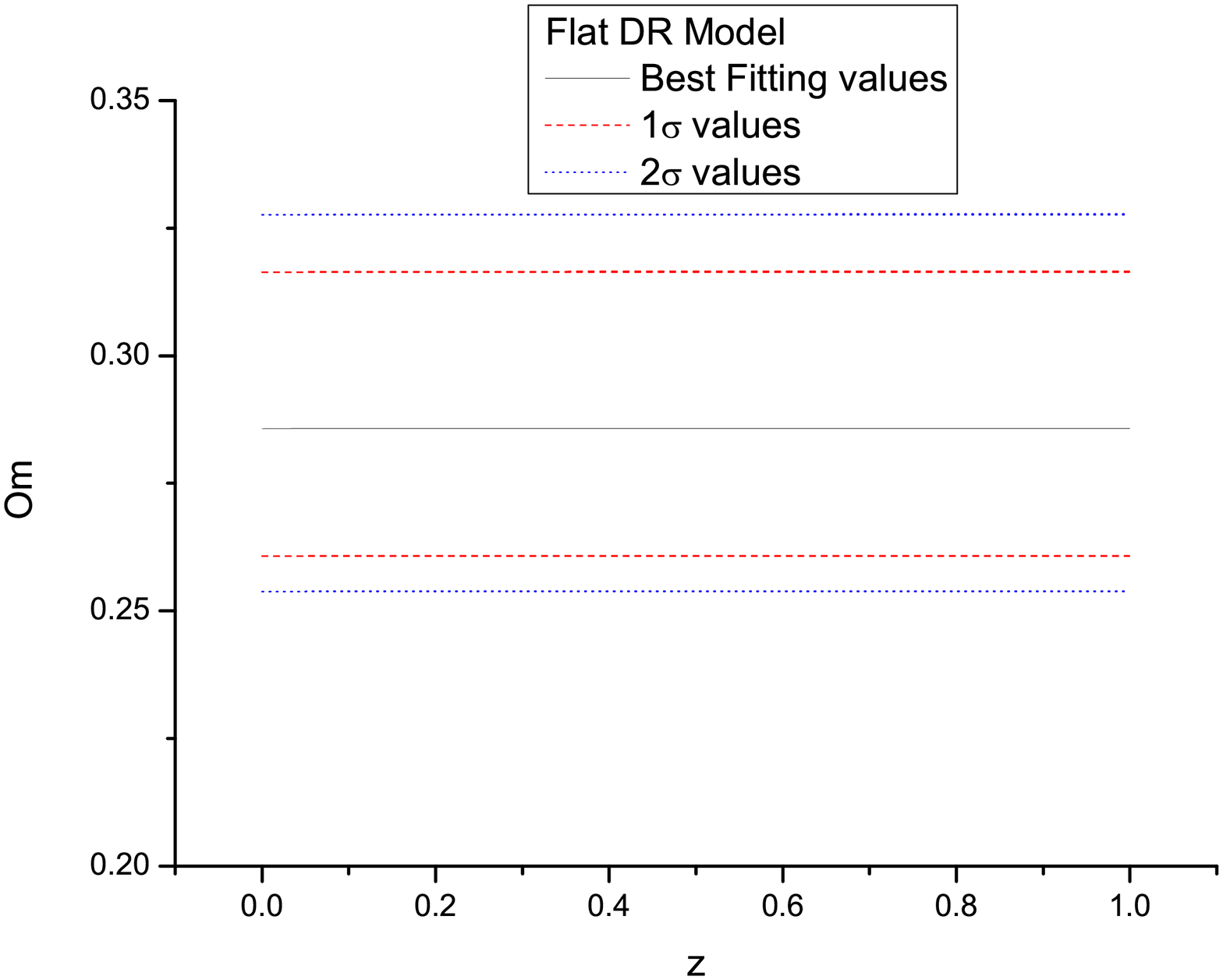}\label{omflat}}\quad
 {\includegraphics[width=2.0in]{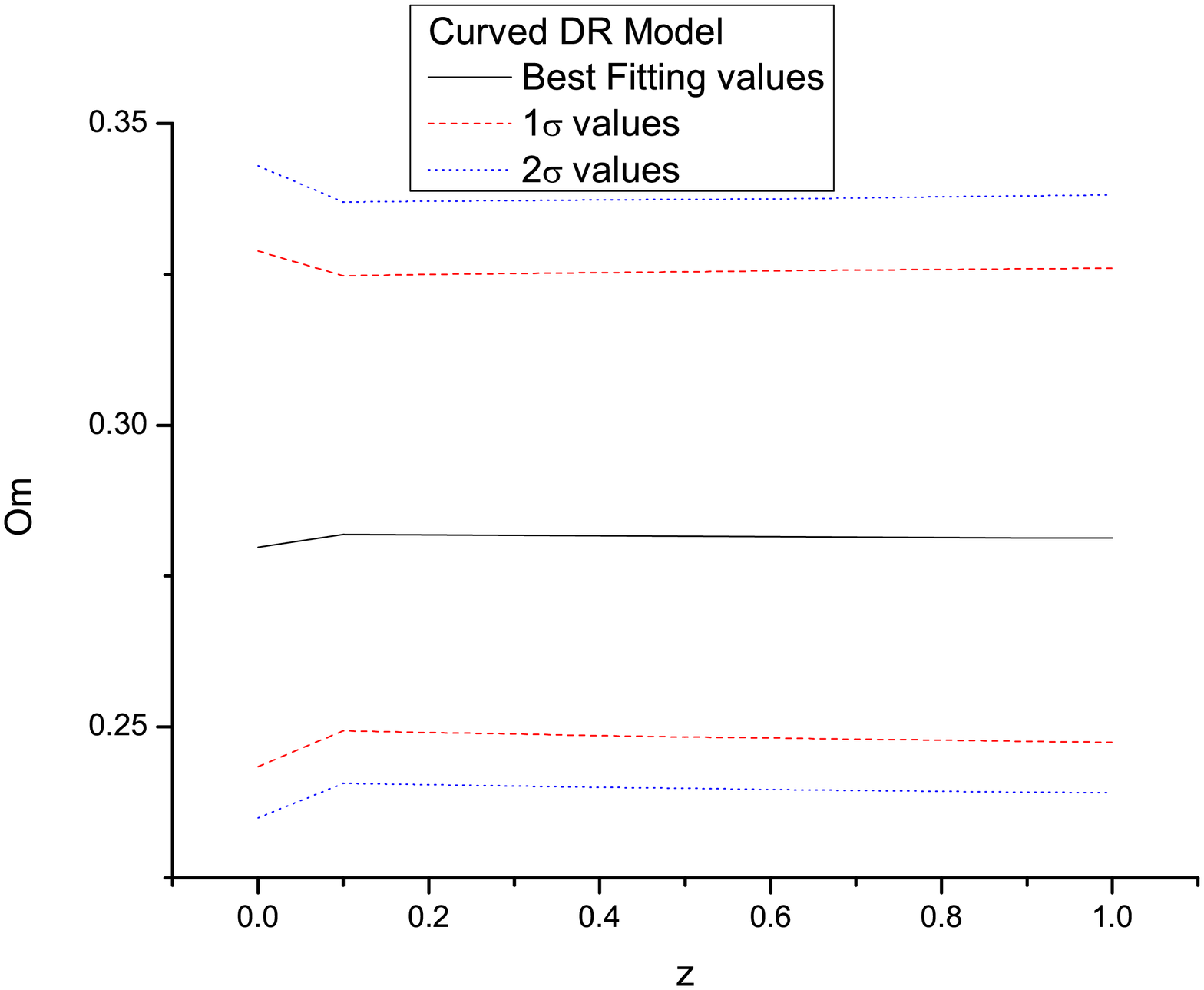}\label{omcurved}}\quad
\caption{ The Om diagnostic for the curved $\Lambda CDM$ model, the flat and curved dark radiation models are give out.
  }\label{Om}
\end{figure}

\begin{figure} \centering
{\includegraphics[width=2.4in]{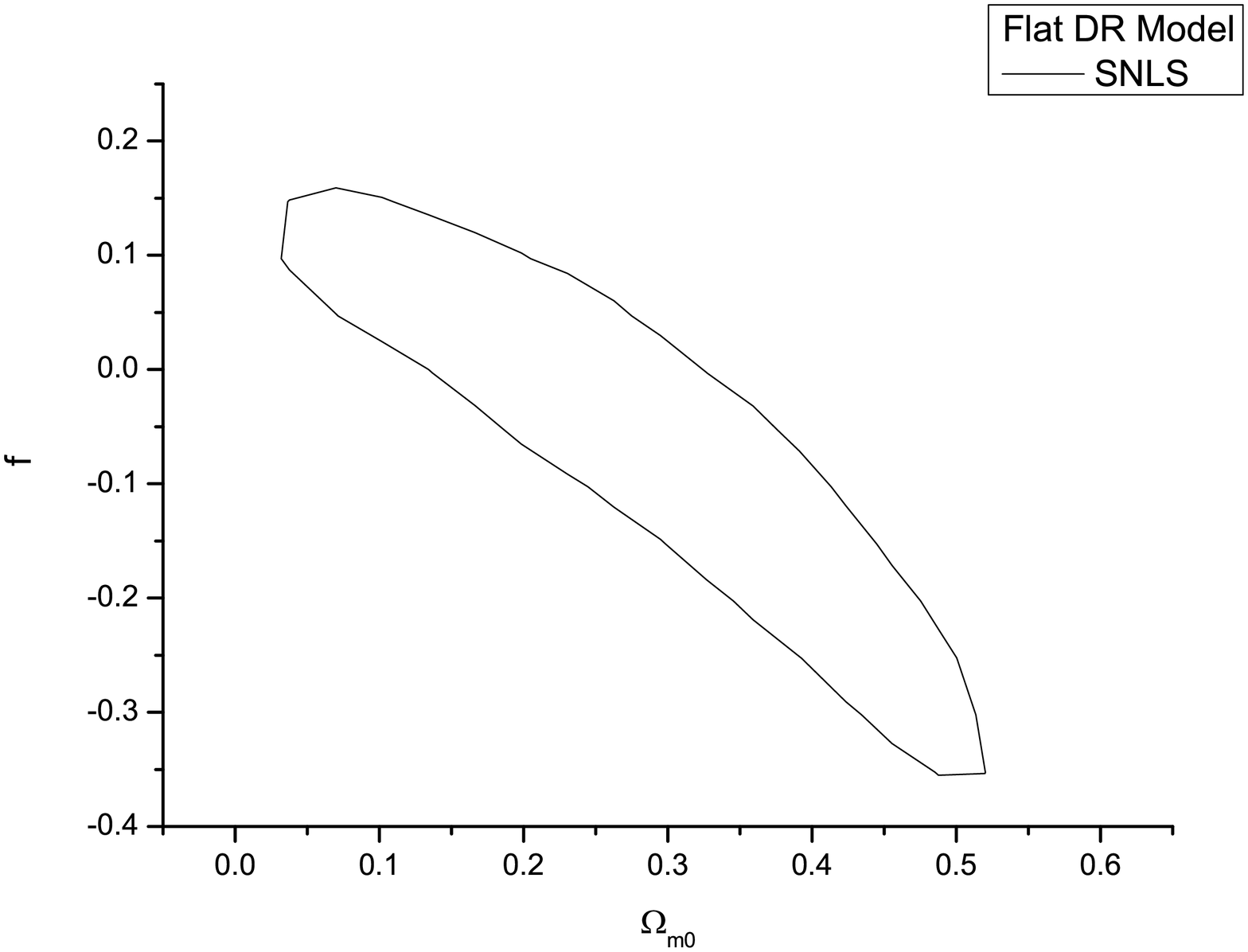}\label{snofm}}\quad
{\includegraphics[width=2.4in]{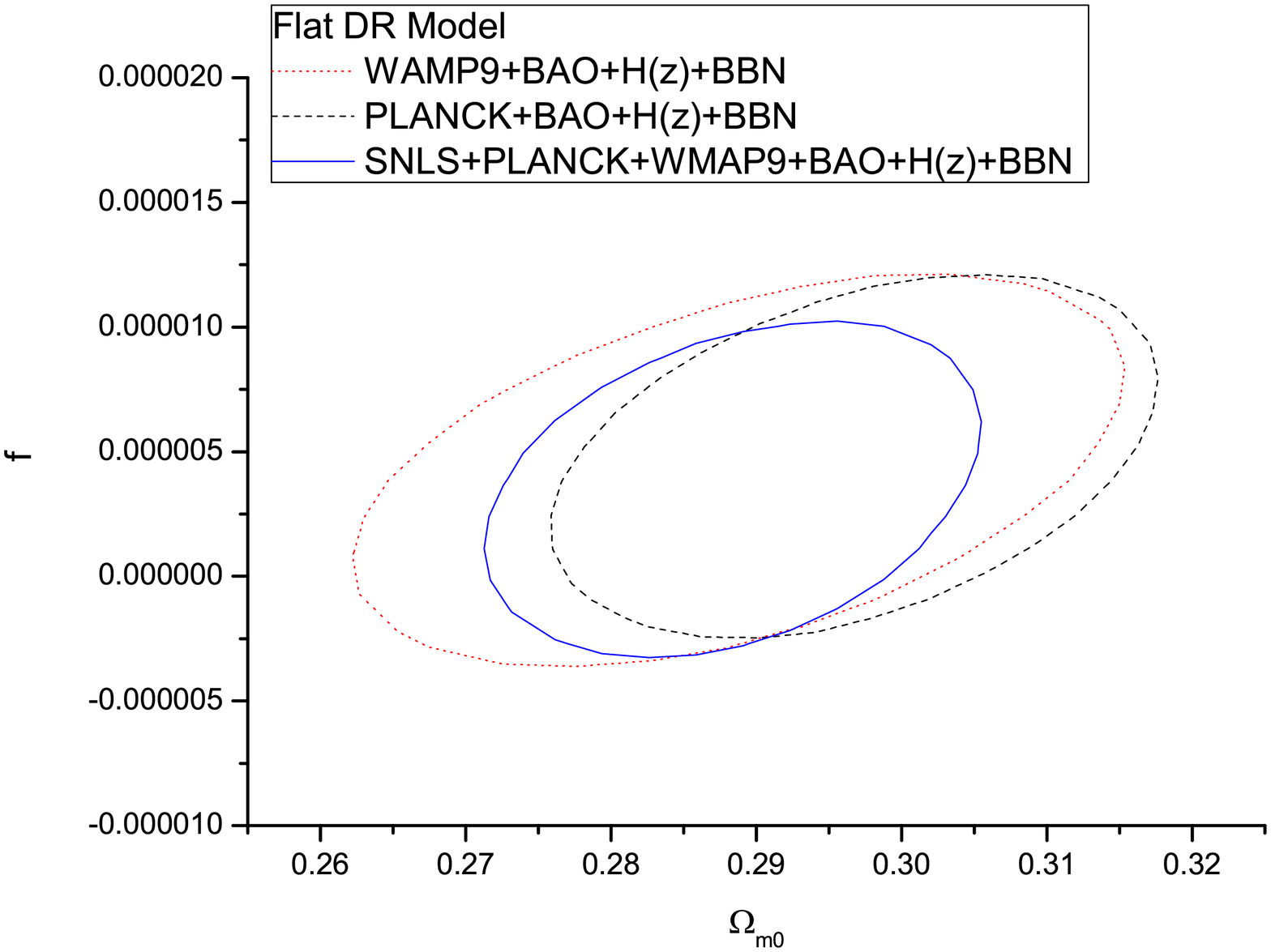}\label{nofm}}\quad
 {\includegraphics[width=2.4in]{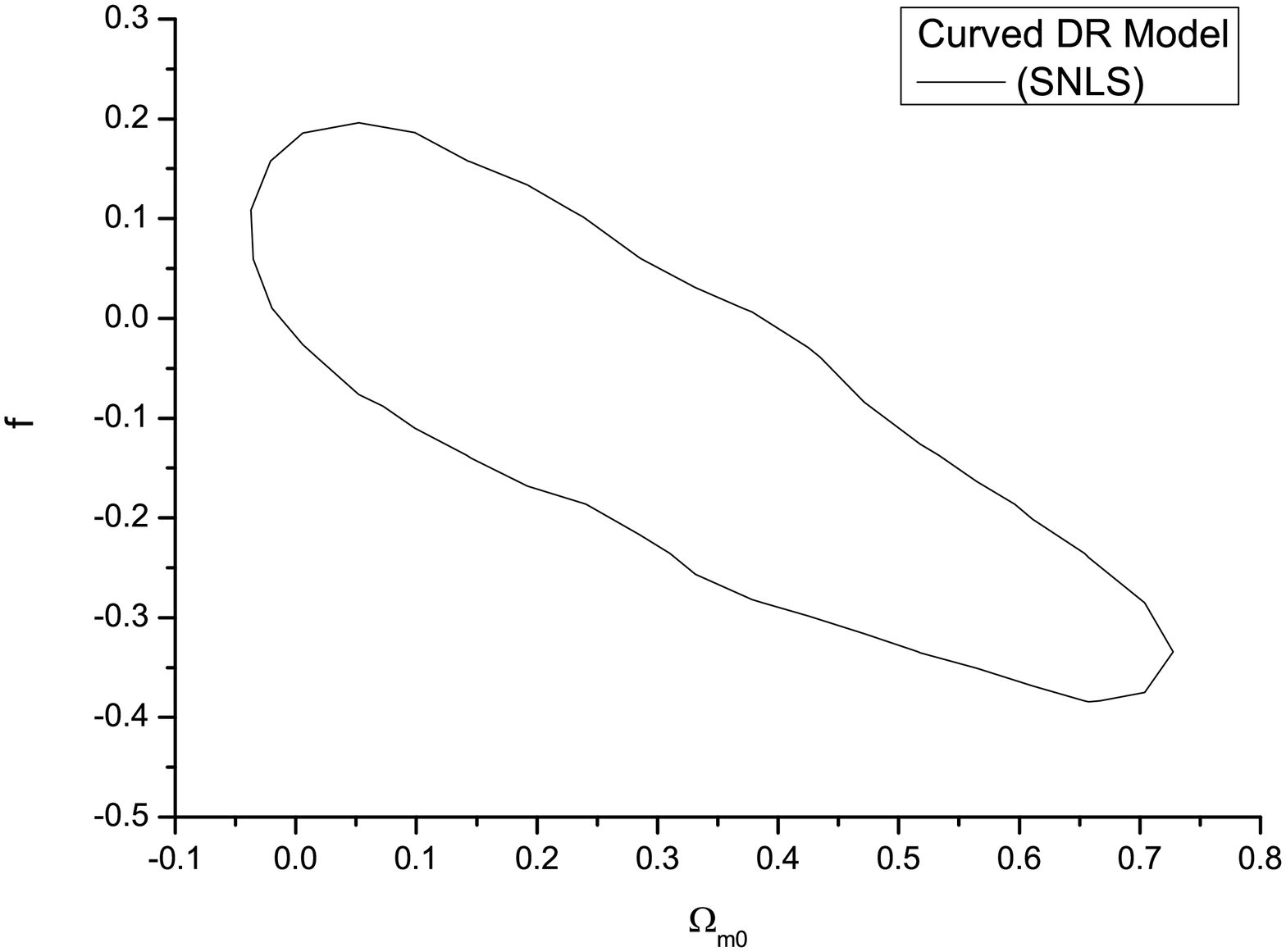}\label{kfm}}\quad
  {\includegraphics[width=2.4in]{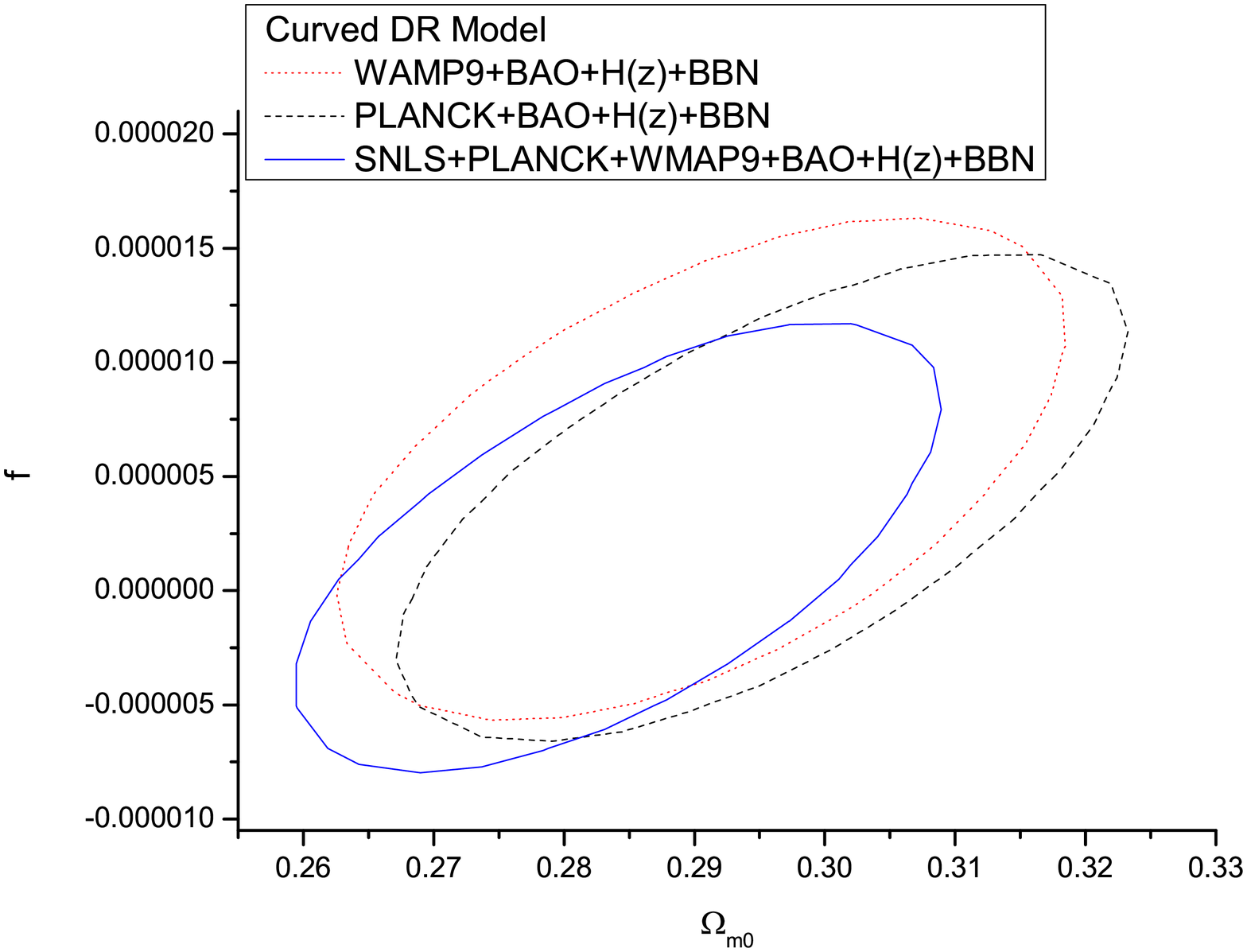}\label{kfm}}\quad
{\includegraphics[width=2.4in]{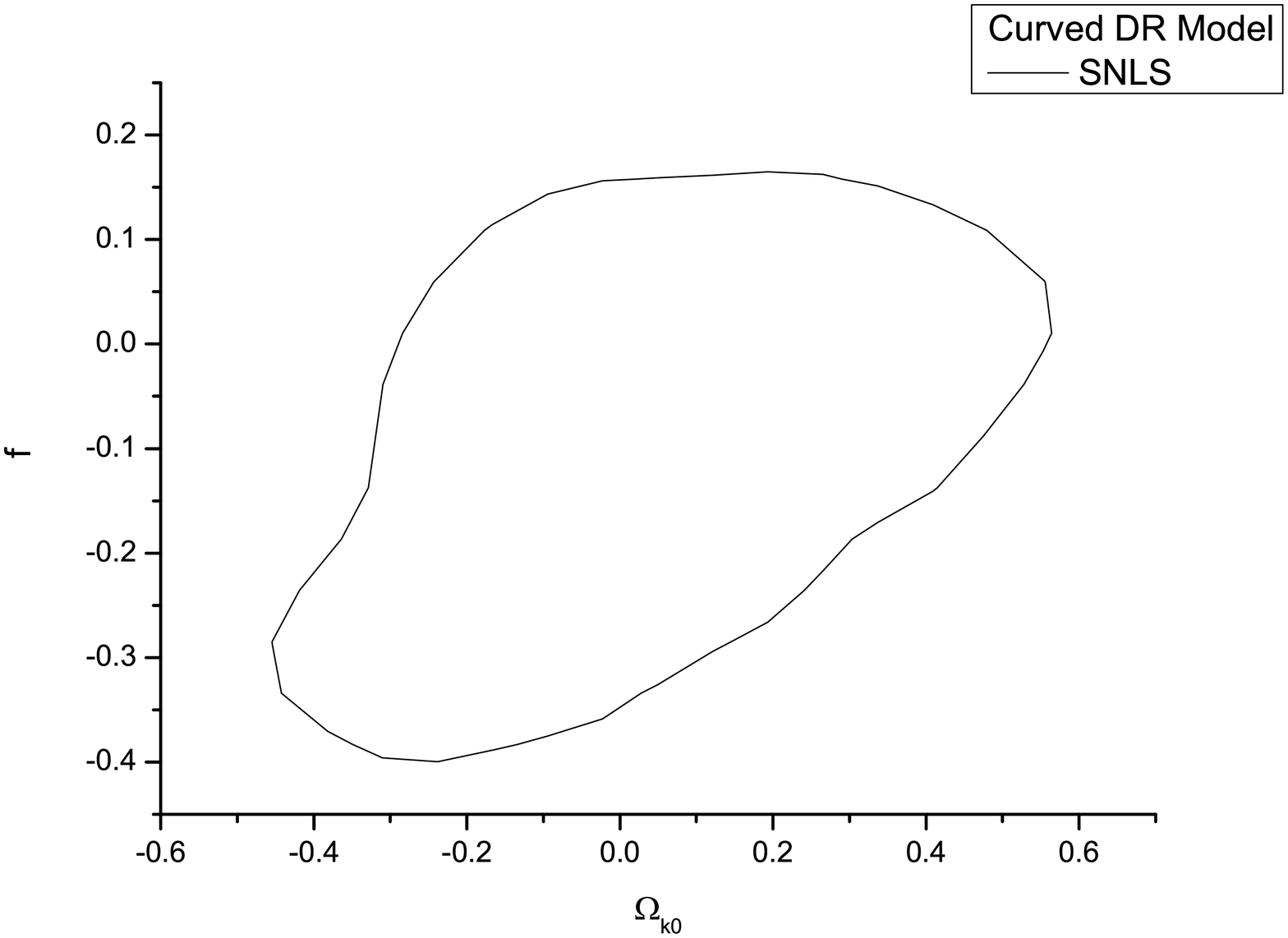}\label{kkf}}\quad
{\includegraphics[width=2.4in]{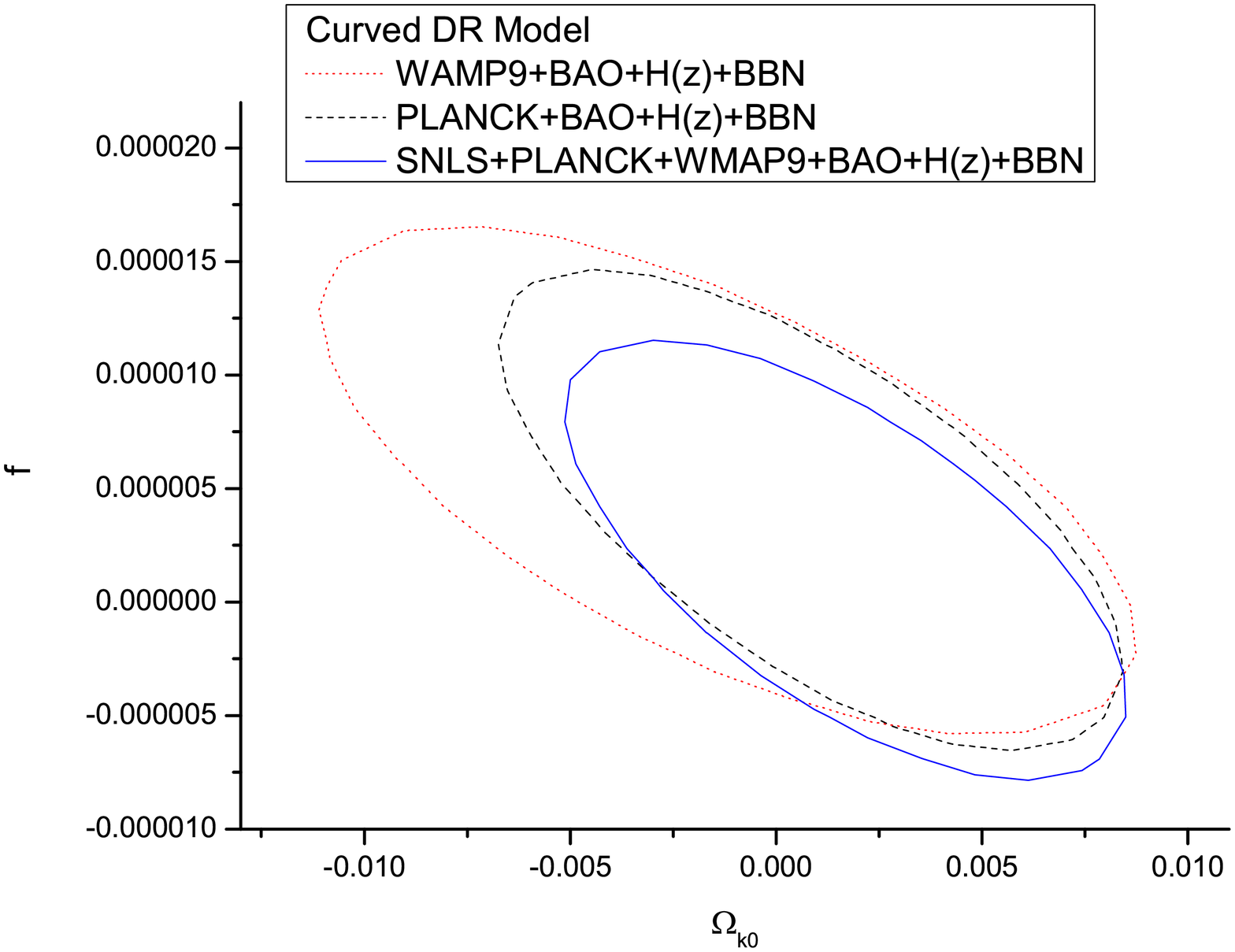}\label{kkf}}\quad
 {\includegraphics[width=2.4in]{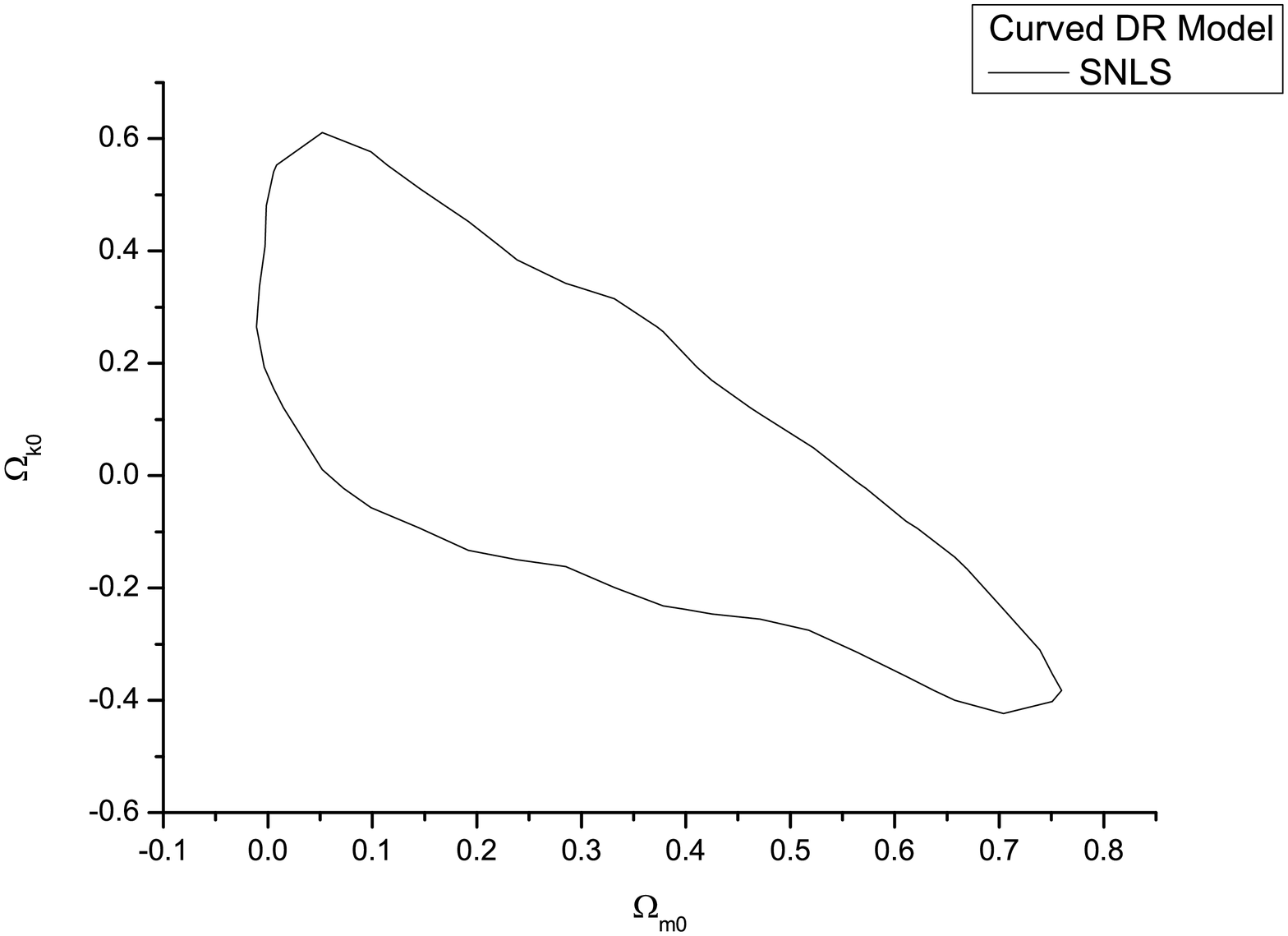}\label{kkm}}\quad
  {\includegraphics[width=2.4in]{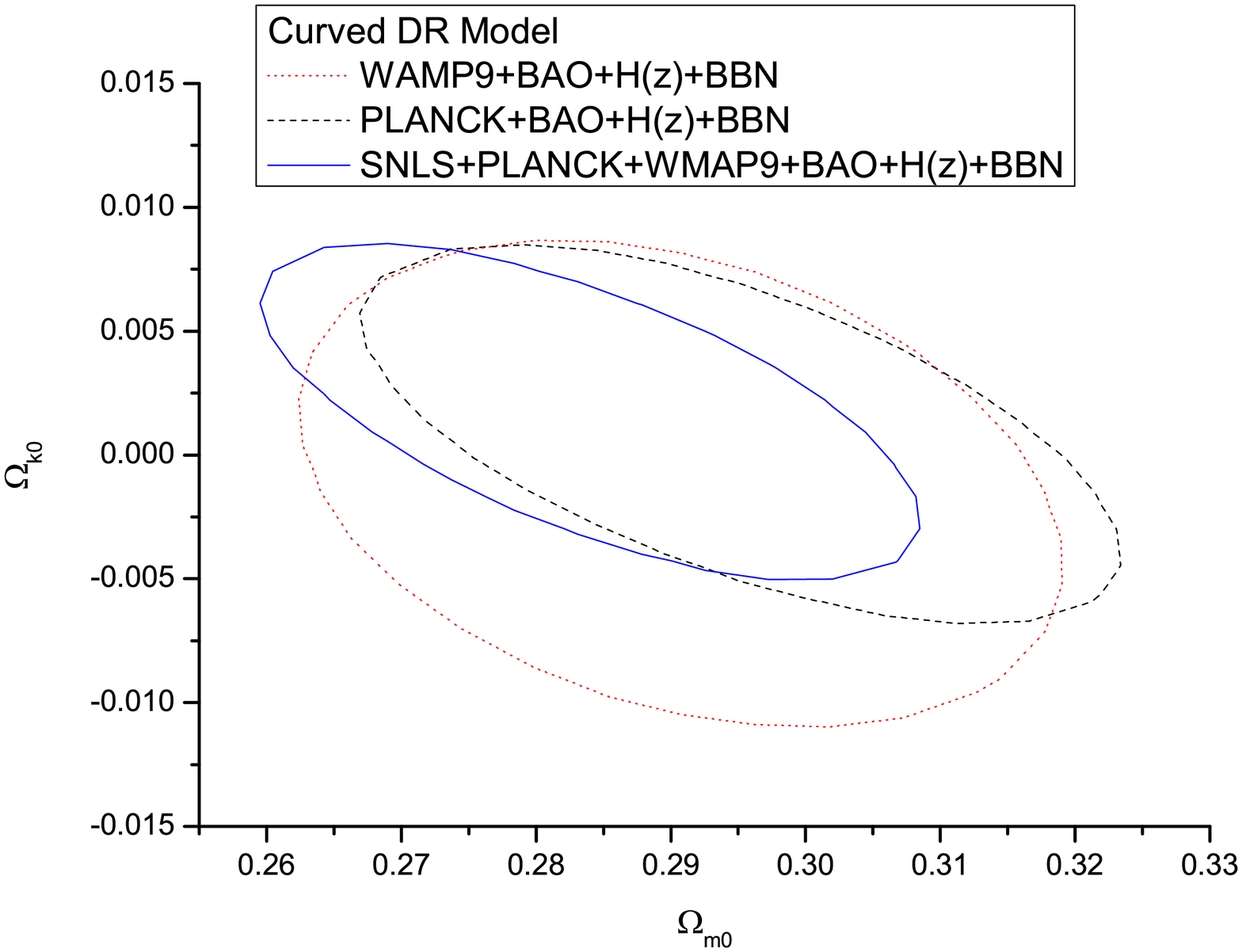}\label{kkm}}\quad
  \caption{The $1\sigma$ parameter contours  given by different data are presented.  }\label{deg}
\end{figure}

\subsubsection{Parameter Degeneration}

 Figure \ref{deg} presents the contours of  $\Omega_{m0}-f$,  $\Omega_{k0}-f$ and $\Omega_{m0}-\Omega_{k0}$  of the flat and curved dark radiation models given by different datasets. As we mentioned above,  the $SNLS$ data  give out loose constraints. Meanwhile, the three data (Data\Rmnum{2},  \Rmnum{3} and \Rmnum{4}) give much tighter constraints which also  have the same contour directions.
 For the contour of $\Omega_{m0}-\Omega_{k0}$,  all the data give the same constrain direction. In contrast,  for  $\Omega_{m0}-f$ and $\Omega_{k0}-f$,   the SNLS data and  the other data  gives  contours with different directions. Obviously,  the degeneration between  the dark radiation parameter $f$ and the other parameters need more data to  break.

\section{The Summary}\label{sec6}
Theoretically, after adding  dark radiation,  the phase-analysis method proved that the universe derived from the dark radiation model   could go through the radiation
dominated phase, the matter dominated phase and the dark energy dominated phase sequently. In a conclusion, the  model  is compatible with the
history of our universe. 

Observationally, we use the  
$SNLS$, $WMAP9$, $PLANCK$, 
$BAO$, $H(z)$ and $BBN$ data to constrain the dark radiation part.
As expected,
 the dark radiation  wiped out the  tension between the $SNLS$ data and the other data in flat $\Lambda CDM$ model.
And,  the constraining results are at a  reasonable level, e.g. $f\sim 10^{-5}$. 
  The small dark radiation parameter $f$ give a small deviation of $\omega_{0}$ and
$\omega_{0}'$.  And, the effect of dark radiation make  the best fit value  of $N_{eff}$  slightly larger than $3.04$. Anyway, the $Om$ diagnostic could extract the curved dark radiation from the flat $\Lambda CDM $ model, but it has no effect on the flat dark radiation model. And,  more data are needed for dark radiation because of parameter degenerations.


\section{Acknowledgements}
We are grateful to the useful suggestions from the anonymous referee. 
YZ thanks the useful discussion with  Dr. Hao Wang, Dr. Hongbo Zhang, Dr. Yu Pan, Prof. Nana Pan.
This work was supported by   the Ministry of Science and Technology of China
national basic science Program (973 Project) under grant Nos. 2010CB833004, the National Natural Science
Foundation of China project under grant Nos. 11175270, 11005164, 11073005
and 10935013,
CQ CSTC under grant
No. 2010BB0408, and CQ MEC under grant No. KJTD201016.  Part of this research was supported under the U.S.
Department of Energy contract DE-AC02-06CH11357 and by the DOE under contract W-7405-ENG-36.


{}

\begin{thebibliography}{}






\bibitem{Riess}
A.G. Riess  et al., AJ. {\bf 116}, 1009  (1998).

\bibitem{Perlmutter99}
S. Perlmutter et
al., ApJ {\bf 517}, 565 (1999).

\bibitem{Tonry03}
J. L. Tonry  et al., ApJ {\bf 594},
1 (2003).

\bibitem{Knop03}
R.A. Knop et al., ApJ {\bf 598}, 102 (2003).

\bibitem{Riess04}
A.G. Riess et
al., ApJ {\bf 607}, 665 (2004).

\bibitem{Bennet03}  C.L. Bennet et al., ApJS. 148, 1 (2003).

\bibitem{spergel03}
D.N. Spergel et al., ApJS, 148, 175 (2003); D.N. Spergel et al.,
ApJS {\bf 170}, 377 (2007).

\bibitem{Page07}
 L. Page et al., ApJS {\bf 170}, 335
(2007).

\bibitem{Hinshaw07}
G. Hinshaw et al., ApJS {\bf 170}, 263 (2007).


\bibitem{Komatsu:2010fb}
  E.~Komatsu {\it et al.}  [WMAP Collaboration],
  Astrophys.\ J.\ Suppl.\  {\bf 192}, 18 (2011)
  [arXiv:1001.4538 [astro-ph.CO]].

\bibitem{Randall:1999ee}
  L.~Randall and R.~Sundrum,
  Phys.\ Rev.\ Lett.\  {\bf 83}, 3370 (1999)
  [arXiv:hep-ph/9905221].

\bibitem{Vishwakarma:2002ek}
  R.~G.~Vishwakarma and P.~Singh,
  Class.\ Quant.\ Grav.\  {\bf 20}, 2033 (2003)
  [arXiv:astro-ph/0211285].
\bibitem{Ichiki:2002eh}
  K.~Ichiki, M.~Yahiro, T.~Kajino, M.~Orito and G.~J.~Mathews,
  Phys.\ Rev.\  D {\bf 66}, 043521 (2002)
  [arXiv:astro-ph/0203272].


\bibitem{Ichiki:2002yp}
  K.~Ichiki, P.~M.~Garnavich, T.~Kajino, G.~J.~Mathews and M.~Yahiro,
  Phys.\ Rev.\  D {\bf 68}, 083518 (2003)
  [arXiv:astro-ph/0210052].

\bibitem{Calabrese:2011hg}
  E.~Calabrese, D.~Huterer, E.~V.~Linder, A.~Melchiorri and L.~Pagano,
  Phys.\ Rev.\  D {\bf 83} (2011) 123504
  [arXiv:1103.4132 [astro-ph.CO]].

\bibitem{DeFelice:2005bx}
  A.~De Felice, G.~Mangano, P.~D.~Serpico and M.~Trodden,
  Phys.\ Rev.\  D {\bf 74}, 103005 (2006)
  [arXiv:astro-ph/0510359].

\bibitem{Dutta:2009jn}
  S.~Dutta and E.~N.~Saridakis,
  JCAP {\bf 1001} (2010) 013
  [arXiv:0911.1435 [hep-th]].

\bibitem{Ali:2011sv} 
  A.~Ali, S.~Dutta, E.~N.~Saridakis and A.~A.~Sen,
  Gen.\ Rel.\ Grav.\  {\bf 44}, 657 (2012)
  [arXiv:1004.2474 [astro-ph.CO]].

\bibitem{Lima:2000ay}
  J.~A.~S.~Lima, A.~I.~Silva and S.~M.~Viegas,
  Mon.\ Not.\ Roy.\ Astron.\ Soc.\  {\bf 312}, 747 (2000).

\bibitem{Birkel:1996py}
  M.~Birkel and S.~Sarkar,
  Astropart.\ Phys.\  {\bf 6}, 197 (1997)
  [arXiv:astro-ph/9605055].

\bibitem{Lima:1995kd}
  J.~A.~S.~Lima,
  Phys.\ Rev.\  D {\bf 54}, 2571 (1996)
  [arXiv:gr-qc/9605055].

\bibitem{Bordag:2001qi}
  M.~Bordag, U.~Mohideen and V.~M.~Mostepanenko,
  Phys.\ Rept.\  {\bf 353}, 1 (2001)
  [arXiv:quant-ph/0106045].

\bibitem{Dunkley:2010ge}
  J.~Dunkley {\it et al.},
  Astrophys.\ J.\  {\bf 739}, 52 (2011)
  [arXiv:1009.0866 [astro-ph.CO]].
  
  
  
  \bibitem{ACT}
S.Das et al.,arXiv:1301.1037; R.Keisler et al.,ApJ 74328(2011); C.L.Reichardt et al.,ApJ 763,127(2013);
26
C. L. Reichardt et al., ApJ 755, 70 (2012); K. T. Story et al., arXiv:1210.7231.


\bibitem{Keisler:2011aw}
  R.~Keisler, C.~L.~Reichardt, K.~A.~Aird, B.~A.~Benson, L.~E.~Bleem, J.~E.~Carlstrom, C.~L.~Chang and H.~M.~Cho {\it et al.},
  Astrophys.\ J.\  {\bf 743}, 28 (2011)
  [arXiv:1105.3182 [astro-ph.CO]].

\bibitem{Cyburt:2004yc}
  R.~H.~Cyburt, B.~D.~Fields, K.~A.~Olive and E.~Skillman,
  Astropart.\ Phys.\  {\bf 23}, 313 (2005)
  [arXiv:astro-ph/0408033].

\bibitem{Fischler:2010xz}
  W.~Fischler and J.~Meyers,
  Phys.\ Rev.\  D {\bf 83}, 063520 (2011)
  [arXiv:1011.3501 [astro-ph.CO]].


\bibitem{Zhang:2012zz}
  Y.~Zhang, H.~Li, Y.~Gong and Z.~-H.~Zhu,
  Eur.\ Phys.\ J.\ C {\bf 72}, 2035 (2012).

\bibitem{Hamann:2011hu}
  J.~Hamann,
  JCAP {\bf 1203}, 021 (2012)
  [arXiv:1110.4271 [astro-ph.CO]].

\bibitem{Menestrina:2011mz}
  J.~L.~Menestrina and R.~J.~Scherrer,
  Phys.\ Rev.\  D {\bf 85}, 047301 (2012)
  [arXiv:1111.0605 [astro-ph.CO]].

\bibitem{Dutta:2009ix}
  S.~Dutta, S.~D.~H.~Hsu, D.~Reeb and R.~J.~Scherrer,
  Phys.\ Rev.\  D {\bf 79} (2009) 103504
  [arXiv:0902.4699 [astro-ph.CO]].

\bibitem{Conley:2011ku}
  A.~Conley {\it et al.},
  Astrophys.\ J.\ Suppl.\  {\bf 192}, 1 (2011)
  [arXiv:1104.1443 [astro-ph.CO]].



\bibitem{Sullivan:2011kv}
  M.~Sullivan {\it et al.},
  Astrophys.\ J.\  {\bf 737}, 102 (2011)
  [arXiv:1104.1444 [astro-ph.CO]].
  
  
  
  
\bibitem{Bennett:2012fp}
  C.~L.~Bennett, D.~Larson, J.~L.~Weiland, N.~Jarosik, G.~Hinshaw, N.~Odegard, K.~M.~Smith and R.~S.~Hill {\it et al.},
  arXiv:1212.5225 [astro-ph.CO].


\bibitem{Hinshaw:2012aka}
  G.~Hinshaw {\it et al.}  [WMAP Collaboration],
  arXiv:1212.5226 [astro-ph.CO].

\bibitem{Ade:2013tyw}
  P.~A.~R.~Ade {\it et al.}  [Planck Collaboration],
  arXiv:1303.5077 [astro-ph.CO].

\bibitem{Percival:2007yw}
  W.~J.~Percival, S.~Cole, D.~J.~Eisenstein, R.~C.~Nichol, J.~A.~Peacock, A.~C.~Pope and A.~S.~Szalay,
  Mon.\ Not.\ Roy.\ Astron.\ Soc.\  {\bf 381}, 1053 (2007)
  [arXiv:0705.3323 [astro-ph]].

\bibitem{Percival:2009xn}
  W.~J.~Percival {\it et al.}  [SDSS Collaboration],
  Mon.\ Not.\ Roy.\ Astron.\ Soc.\  {\bf 401}, 2148 (2010)
  [arXiv:0907.1660 [astro-ph.CO]].

\bibitem{Blake:2011en}
  C.~Blake {\it et al.},
  Mon.\ Not.\ Roy.\ Astron.\ Soc.\  {\bf 418}, 1707 (2011)
  [arXiv:1108.2635 [astro-ph.CO]].

\bibitem{Beutler:2011hx}
  F.~Beutler {\it et al.},
  Mon.\ Not.\ Roy.\ Astron.\ Soc.\  {\bf 416}, 3017 (2011)
  [arXiv:1106.3366 [astro-ph.CO]].


\bibitem{Simon:2004tf}
  J.~Simon, L.~Verde and R.~Jimenez,
  Phys.\ Rev.\  D {\bf 71}, 123001 (2005)
  [arXiv:astro-ph/0412269].

\bibitem{Riess:2009pu}
  A.~G.~Riess {\it et al.},
  Astrophys.\ J.\  {\bf 699}, 539 (2009)
  [arXiv:0905.0695 [astro-ph.CO]].

\bibitem{Serra:2009yp}
  P.~Serra, A.~Cooray, D.~E.~Holz, A.~Melchiorri, S.~Pandolfi and D.~Sarkar,
  Phys.\ Rev.\  D {\bf 80}, 121302 (2009)
  [arXiv:0908.3186 [astro-ph.CO]].

\bibitem{Burles:2000zk}
  S.~Burles, K.~M.~Nollett and M.~S.~Turner,
  Astrophys.\ J.\  {\bf 552}, L1 (2001)
  [arXiv:astro-ph/0010171].

\bibitem{Steigman:2005uz}
  G.~Steigman,
  Int.\ J.\ Mod.\ Phys.\  E {\bf 15}, 1 (2006)
  [arXiv:astro-ph/0511534].

    \bibitem{dynamics}
A.~A.~Coley, gr-qc/9910074.

 \bibitem{Wainwright}
J.~Wainwright and G.~F.~R.~Ellis, {\it Dynamical Systems in Cosmology}, Cambridge Univ. Press, Cambridge, 1997.

   \bibitem{Coley}
A.~A.~Coley, {\it Dynamical Systems and Cosmology}, in Series: Astrophysics and Space Science Library, Vol.~291, Springer, 2004.


\bibitem{Eisenstein:1997ik}
  D.~J.~Eisenstein and W.~Hu,
  Astrophys.\ J.\  {\bf 496}, 605 (1998)
  [arXiv:astro-ph/9709112].

\bibitem{Hui:2005nm}
   L.~Hui and P.~B.~Greene,
  Phys.\ Rev.\  D {\bf 73}, 123526 (2006)
  [arXiv:astro-ph/0512159].

\bibitem{Hu:1995en}
  W.~Hu and N.~Sugiyama,
  Astrophys.\ J.\  {\bf 471}, 542 (1996)
  [arXiv:astro-ph/9510117].



  \bibitem{Sahni:2008xx}
  V.~Sahni, A.~Shafieloo and A.~A.~Starobinsky,
  Phys.\ Rev.\ D {\bf 78}, 103502 (2008)
  [arXiv:0807.3548 [astro-ph]].

\bibitem{Shafieloo:2012rs}
  A.~Shafieloo, V.~Sahni and A.~A.~Starobinsky,
  Phys.\ Rev.\ D {\bf 86} (2012) 103527
  [arXiv:1205.2870 [astro-ph.CO]].


\bibitem{Godlowski:2006vf}
  W.~Godlowski and M.~Szydlowski,
  Phys.\ Lett.\  B {\bf 642}, 13 (2006)
  [arXiv:astro-ph/0606731].

\bibitem{Mangano:2006ur}
  G.~Mangano, A.~Melchiorri, O.~Mena, G.~Miele and A.~Slosar,
  JCAP {\bf 0703}, 006 (2007)
  [arXiv:astro-ph/0612150].

\bibitem{Bashinsky:2003tk}
  S.~Bashinsky and U.~Seljak,
  Phys.\ Rev.\  D {\bf 69}, 083002 (2004)
  [arXiv:astro-ph/0310198].

\bibitem{Archidiacono:2011gq} 
  M.~Archidiacono, E.~Calabrese and A.~Melchiorri,
  Phys.\ Rev.\ D {\bf 84}, 123008 (2011)
  [arXiv:1109.2767 [astro-ph.CO]].

  \bibitem{Clarkson:2007bc}
  C.~Clarkson, M.~Cortes and B.~A.~Bassett,
  JCAP {\bf 0708}, 011 (2007)
  [astro-ph/0702670 [ASTRO-PH]].

\bibitem{Okouma:2012dy} 
  P.~M.~Okouma, Y.~Fantaye and B.~A.~Bassett,
  Phys.\ Lett.\ B {\bf 719}, 1 (2013)
  [arXiv:1207.3000 [astro-ph.CO]].

\bibitem{Wang:2011sb}
  Y.~Wang, C.~-H.~Chuang and P.~Mukherjee,
  Phys.\ Rev.\ D {\bf 85}, 023517 (2012)
  [arXiv:1109.3172 [astro-ph.CO]].

\bibitem{Wang:2013mha}
  Y.~Wang and S.~Wang,
  arXiv:1304.4514 [astro-ph.CO].



\bibitem{Akaike}
  H.~Akaike,
  IEEE Trans.\ Auto.\ Control.  {\bf 19}, 716 (1974).

\bibitem{Liddle:2004nh}
  A.~R.~Liddle,
  Mon.\ Not.\ Roy.\ Astron.\ Soc.\  {\bf 351}, L49 (2004)
  [arXiv:astro-ph/0401198].




\end{thebibliography}
\end{document}